\documentstyle[prl,aps,floats,epsf,color]{revtex}
\begin{document}
\baselineskip=12pt
\def\be{\begin{equation}}
\def\ee{\end{equation}}
\def\bea{\begin{eqnarray}}
\def\eea{\end{eqnarray}}
\def\orc{\Omega_{r_c}}
\def\om{\Omega_{\text{m}}}
\def\E{{\rm e}}
\def\bearst{\begin{eqnarray*}}
\def\eearst{\end{eqnarray*}}
\def\peleven{\parbox{11cm}}
\def\peffec{\peight{\bearst\eearst}\hfill\peleven}
\def\pspace{\peight{\bearst\eearst}\hfill}
\def\ptwelve{\parbox{12cm}}
\def\peight{\parbox{8mm}}
\twocolumn[\hsize\textwidth\columnwidth\hsize\csname@twocolumnfalse\endcsname

\title{Observational tests of a two parameter power-law class modified gravity in Palatini formalism}

\author{ Shant Baghram$^{1}$ ,M. Sadegh Movahed$^{2,3}$ and Sohrab Rahvar$^{1}$}
\address{$^{1}$Department of Physics, Sharif University of
Technology, P.O.Box 11365--9161, Tehran, Iran}
\address{$^{2}$Department of Physics, Shahid Beheshti university, G.C., Evin, Tehran
19839, Iran}
\address{$^{3}$ School of Astronomy and Astrophysics,\\
 Institute for Research in Fundamental Sciences (IPM), P. O. Box 19395-5531, Tehran, Iran}

\vskip 1cm

 \maketitle
\begin{abstract}
{\bf{CONTEXT}}: In this work we propose a modified gravity action
$f(R)=(R^n-R^n_{0})^{1/n}$ with two free parameters of $n$ and
$R_{0}$ and derive the dynamics of a universe for this action in
the Palatini formalism.\\
{\bf {AIM}}: We do a cosmological comparison of this model with
observed data to find the best parameters of a  model in a flat universe.\\
{\bf{METHOD}}: To constrain the free parameters of model we use SNIa
type Ia data in two sets of gold and union samples,
CMB-shift parameter, baryon acoustic oscillation, gas mass fraction in cluster of galaxies, and large-scale structure data. \\
{\bf {RESULT}}: The best fit from the observational data results in
the parameters of model in the range of $n=0.98^{+0.08}_{-0.08}$ and
$\Omega_M = 0.25_{+0.1}^{-0.1}$ with one sigma level of confidence
where a standard $\Lambda$CDM universe resides
in this range of solution. \\

\end{abstract}

\newpage
]

\section{Introduction}
Recent observation of CMB+SNIa reveals that the Universe is under
positive acceleration, in contrast to our expectations from the
behavior of ordinary matter. One of the possible solutions is
assuming a cosmological constant to provide a late time acceleration
to the universe \cite{lambda}. Although $\Lambda$CDM is the easiest
model that fits well with the current observational data,it suffers
from the fine-tuning and coincidence problems, which motivates to
introduce alternative theories such as dark energy models. The other
possibility is the modification of gravity law in such a way that
behaves as standard general relativity in a strong gravitational
regime, but repulses particles in the cosmological scales
\cite{modgrv}. Each proposed model must challenge two different
observational criteria of (a) cosmological tests and (b) local
gravity tests in solar system scales.

In this work we examine a $f(R)=(R^n-R^n_{0})^{1/n}$ gravity model
with the cosmological observational data to fix the two parameters
of the model, $R_0$ and $n$. We use Supernova Type Ia, CMB shift
parameter, baryonic acoustic oscillation, gas mass fraction in
cluster of galaxies and large scale structures to find the best
parameter of the model. The organization of paper is as follows: In
Sec. \ref{section2}, we introduce the field equation in $f(R)$
modified gravity model in Palatini formalism. Using FRW metric we
derive a modified Friedman equation. In Sec. \ref{section3} we apply
our proposed action to derive the field equations. In Sec.
\ref{section4} we study geometrical and dynamical behavior of
universe in this model. In Sec. \ref{section5} using the geometrical
observations as SNIa, CMB-shift parameter, baryon acoustic
oscillation, gas mass fraction of clusters of galaxies, we put
constraints on the parameters of the model. Finally, we do
comparison of models with the large scale structures' data in Sec.
\ref{section5}. The conclusions are given in Sec.\ref{section6}. We
show that the best parameter of this model privilege a $\Lambda$CDM
universe.

\section{modified gravity in Palatini formalism}
\label{section2} For an arbitrary action of the gravity as a
function of Ricci scalar $f(R)$, there are two main approaches to
extract the field equations. The first one is the so-called ''metric
formalism'', which is obtained by the variation of action  with
respect to the metric. In this formalism in contrast to the
Einstein-Hilbert action, the field equation is a fourth order
nonlinear differential equation. In the second approach so-called
Palatini formalism, the connection and metric are considered
independent fields and variation of action with respect to these
fields results in a set of second order differential equations. In
what follows we will work in Palatini formalism. Let us take the
general form of action in the Palatini formalism as
\begin{equation}
S[f;g,\hat{\Gamma},\Psi_{m}] =
\frac{1}{2\kappa}\int{d^{4}x\sqrt{-g}f(R)+S_{m}[g_{\mu\nu},\Psi_{m}]},
\end{equation}
where $\kappa=8\pi{G}$ and $S_{m}[g_{\mu\nu},\Psi_{m}]$ is the
matter action that depends on metric $g_{\mu\nu}$ and  the matter
fields $\Psi_{m}$.
$R=R(g,\hat{\Gamma})=g^{\mu\nu}{R_{\mu\nu}}(\hat{\Gamma})$ is the
generalized Ricci scalar and $R_{\mu\nu}$ is the Ricci tensor, made
of affine connection. Varying  action with respect to the metric
results in
\begin{equation}
f'(R)R_{\mu\nu}(\hat{\Gamma})-\frac{1}{2}f(R)g_{\mu\nu}=\kappa
T_{\mu\nu}, \label{field}
\end{equation}
where prime is the differential with respect to the Ricci scalar and
$T_{\mu\nu}$ is the energy-momentum tensor
\begin{equation}
T_{\mu\nu}=\frac{-2}{\sqrt{-g}}\frac{\delta S_{m}}{\delta
g^{\mu\nu}}.
\end{equation}
On the other hand varying the action with respect to the  connection
results in
\begin {equation} \label{connection}
\hat{\nabla_{\alpha}}[f'(R)\sqrt{-g}g^{\mu\nu}]=0,
\end {equation}
where $\hat{\nabla}$ is the covariant derivative defined from
parallel transformation and  depends on  affine connection. From
Eq.(\ref{connection}), we can define a new metric of
$h_{\mu\nu}=f'(R)g_{\mu\nu}$ conformally related to the physical
metric where the connection is the Christoffel symbol of this new
metric. We take a flat FRW metric (namely $K=0$) for the universe
\begin{equation}
ds^{2}=-dt^{2}+a(t)^{2}\delta_{ij}dx^{i}dx^{j},
\end{equation}
and assume that universe is filled with a perfect fluid with the
energy-momentum tensor of $T^{\nu}_{\mu}=diag(-\rho,p,p,p)$. Using
the metric and energy momentum tensor in Eq.(\ref{field}) we obtain
the generalized FRW equations. It should be noted that the
conservation law of energy-momentum tensor, $T^{\mu\nu}{}_{;\mu}=0$
is defined according to the covariant derivative with respect to the
metric to guarantee the motion of particles on geodesics
\cite{gr-qc/0505128}. A combination of $G^0_0$ and $G^i_i$ results
in
\begin{equation}\label{Hub1}
(H+\frac{1}{2}\frac{\dot{f'}}{f'})^{2}=\frac{1}{6}\frac{\kappa(\rho+3p)}{f'}+\frac{1}{6}\frac{f}{f'}.
\label{hp}
\end{equation}
On the other hand, the trace of Eq. (\ref{field}) gives,
\begin{equation}
Rf'(R)-2f(R)=\kappa T, \label{trace}
\end{equation}
where $T=g^{\mu\nu}T_{\mu\nu}=-\rho + 3 p$. The time derivative of
this equation results in $\dot{R}$ in terms of the time derivative
of density and pressure. Using the equation of state of cosmic fluid
$p=p(\rho)$ and continuity equation, the time derivative of Ricci is
obtained as
\begin{equation}
\dot{R}=3\kappa H\frac{(1-3dp/d\rho)(\rho+p)}{Rf''-f'(R)}.
\label{rdot}
\end{equation}
To obtain a generalized first FRW equation, we start with Eq.
(\ref{trace}) and obtain the density of matter in terms of the Ricci
scalar as
\begin{equation} \label{den}
\kappa\rho=\frac{2f-Rf'}{1-3\omega},
\end{equation}
where $w=p/\rho$. Substituting Eq. (\ref{den}) in (\ref{hp}) and
using Eq. (\ref{rdot}) to change $d/dt = \dot{R}d/dR$, we obtain the
dynamics of the universe in terms of the Ricci scalar as
\begin{equation}
H^{2}=\frac{1}{6(1-3\omega)f'}\frac{3(1+\omega)f-(1+3\omega)Rf'}{\left[1+\frac{3}{2}(1+\omega)\frac{f''(2f-Rf')}{f'(Rf''-f')}\right]^{2}}.
\label{hpala}
\end{equation}
On the other hand using Eq. ({\ref{trace}) and the continuity
equation, the scale factor can be obtained in terms of the Ricci
scalar
\begin{equation}
a=\left[\frac{1}{\kappa\rho_{0}(1-3\omega)}(2f-Rf')\right]^{-\frac{1}{3(1+\omega)}},
\label{apala}
\end{equation}
where $\rho_{0}$ is the energy density at the present time and
$a_{0}$, the scale factor at the present time, is set to $1$. Now
for a generic modified action, eliminating the Ricci scalar in favor
of the scale factor between Eqs. (\ref{hpala}) and (\ref{apala}) we
can obtain the dynamics of universe [i.e. $H=H(a)$]. For the simple
case of matter dominant epoch $\omega=0$, these equations reduce to
\begin{equation} \label{Hub}
H^{2}=\frac{1}{6f'}\frac{3f-Rf'}{\left[1+\frac{3}{2}\frac{f''(2f-Rf')}{f'(Rf''-f{'})}\right]^{2}},
\end{equation}
and
\begin{equation} \label{scalefactor}
a=\left[\frac{1}{\kappa\rho_{0}}(2f-R f')\right]^{-\frac{1}{3}}.
\end{equation}

\section{$f(R)=({R^{n}-R_{0}^{n}})^{1/n}$ gravity}
\label{section3} Here in this section we propose a modified gravity
action of $f(R)=(R^n-{R_{0}}^n)^{1/n}$ with the two free parameters
where $R_{0}>0$ and $n>0$. This action is a generalized form of
$n=2$ that has been discussed in \cite{bagh07,bagh07b}. This action
has a minimum vacuum in an empty universe and a flat Minkowski space
is not achievable in this action. This behavior causes an
accelerating expansion of the universe for a low density universe.
The minimum curvature from the vacuum solution in Eq. (\ref{trace})
is:
\begin{equation} \label{vacuum}
R_v=2^{1/n}R_{0}.
\end{equation}
On the other hand, for the strong gravitational regimes the action
reduces to the Einstein-Hilbert action. To have the asymptotic
behavior of action for these two extreme cases we do a Taylor
expansion of action around $R_v$ in an almost empty universe and
$R_v/R \rightarrow 0$ at strong gravitational regimes. For the weak
field, the expansion of action results in
\begin{eqnarray}
f(R) &=& R_v(\frac{1}{2})^{\frac1n} + (\frac{1}{2})^{\frac1n-1}(R-R_v)\nonumber\\
 &+&
(\frac{1}{2})^{\frac1n}\frac{n+1}{R_v}(R-R_v)^2 + ...
\end{eqnarray}
where ignoring higher order terms, we can rewrite this equation as
\begin{equation}
f(R) = R - \Lambda(R_v,n).
\end{equation}
Here $\Lambda(R_v,n)$ is an effective cosmological constant depends
on the curvature in a vacuum and the exponent of action. On the
other hand we expand the action in a strong gravitational regime
(e.g. $R \gg R_v$). In this case the action can be written as
follows:
\begin{equation}
f(R) = R + \sum_{m=1}^{\infty} \frac{1}{m!}\prod_{k =
0}^{m-1}(\frac1n - k)(-1)^m(\frac{R_0}{R})^{mn}R.
\end{equation}
Ignoring the higher orders in a strong gravitational field, this
action reduce to the Einstein-Hilbert action. So our chosen actions
in these two extreme regimes vary from the Einstein-Hilbert to the
Einstein-Hilbert plus cosmological constant.

Let us study the solutions of modified gravity in three common cases
of a pointlike source in vacuum space, a universe in radiation, and
matter-dominant epochs. For a pointlike source in an empty space,
letting $T^{\mu\nu} = 0$ outside the star, we will have a constant
Ricci scalar. This means that we will have constant $R_v$, $f(R_v)$
and $f'(R_v)$ for all the space. With this condition we write the
field equation as follows:
\begin{equation}
G_{\mu\nu} = -\frac12(R_v - \frac{f(R_v)}{f'(R_v)})g_{\mu\nu},
\end{equation}
where the coefficient of metric at the right-hand side of the
equation plays the role of effective cosmological constant.
Substituting the corresponding value for the vacuum from Eq.
(\ref{vacuum}), the effective cosmological constant obtain
$\Lambda_{eff} = 2^{\frac{1}{n}-2}R_0$. So the solution of the field
equation in the spherically symmetric space results in a
Schwarzschild-de-Sitter space. The value of $R_0$ will be fixed in
the next section from the cosmological observations.

In the radiation dominated era we have $p=\rho/3$. With this
equation of state, the trace of the energy-momentum tensor is zero
and it resembles a vacuum solution where the Ricci scalar is
constant and equal to $2^{1/n}R_{0}$. We substitute $f(R)$ and its
derivatives in Eq.(\ref{Hub1}) to have the dynamics of the Hubble
parameter as a function of density of the universe
\begin{equation}
H^2=\frac{2^{\frac{1}{n}-2}}{3}(2\kappa\rho + R_0).
\end{equation}
Analysis for $n=2$ shows that $R_0$ is in the order of $H_0^2$
\cite{bagh07,bagh07b}. So for the radiation-dominant epoch, we
neglect $R_0$ in comparison with the density of the universe. Using
the continuity equation provides $\rho\propto a^{-4}$, then the
scale factor changes with time as $a\propto t^{1/2}$. This result
shows no dynamical deviation from the standard cosmology at the
early universe.

For the matter-dominant epoch, we calculate the dynamics of the
universe for simplicity in terms of a new variable,
$X\equiv{R}/{H_{0}^2}$.The action can be written in this new form as
\begin{eqnarray}
f(R)&=&{H_{0}}^2F(X)\\
F(X)&=&(X^n-{X_{0}}^n)^{1/n}
\end{eqnarray}
where $X_{0}\equiv{R_{0}}/{H_{0}^2}$ and $H_0=100h$ Km/s/Mpc. The
relation between the derivatives with respect to $R$ and new
variable $X$ is related as
\begin{eqnarray}
f'(R)=F'(X),\\
f''(R)=\frac{F''(X)}{H_0^2},
\end{eqnarray}
where the  derivatives in the left-hand side of equations are with
respect to the Ricci scalar, but in the right-hand side they are  in
terms of $X$ , (i.e. $'=\frac{d}{dX}$).

We rewrite Eq. (\ref{Hub}) with the new dimensionless parameter X:
\begin{equation}
{\cal{H}}^2(X)=\frac{1}{6F'}\frac{3F-XF'}{(1+\frac{3}{2}\frac{F''(2F-XF')}{F'(XF''-F')})^{2}},
\label{hcal}
\end{equation}
where ${\cal{H}}(X)= H/H_0$ is the normalized Hubble parameter to
its current value. Using the conventional definition of $\Omega_{m}$
at the present time as
$\Omega_{m}={\kappa\rho^{(0)}_{m}}/{3H_0^{2}}$ and Eq. (\ref{den})
we obtain
\begin{equation}  \label{omega}
\Omega_{m}(X)=\frac{2F-XF'}{3},
\end{equation}
where $\Omega_{m}(X)=\Omega_{m} a^{-3}$. We obtain the relation
between the scale factor and dimensionless parameter $X$ as:
\begin{equation} \label{scale}
a=(\frac{2F-XF'}{3\Omega_{m}})^{-1/3}.
\end{equation}
In order to have positive scale factor, $X$ should change in the
range of $X\ge 2^{\frac{1}{n}}X_{0}$, where the minimum value for
$X_0$ is in agreement with the vacuum solution of the Ricci scalar.

An important point worth  mentioning here is that the two free
parameters $X_{0}$ and $n$, appeared in the dynamics of the universe
can be replaced with more relevant ones. One of them is $\Omega_{m}$
presented in Eq.(\ref{omega}), where depends directly to
$n$, $X_0$ and $X_p$ (p stands for the present time). $X_p$ can be
eliminated using Eq. (\ref{hcal}), letting ${\cal{H}}(X_p) = 1$
results in a relation between $X_P$ and $X_0$ and $n$. The second
parameter we will use instead of $X_0$ is $X_p$.

\section{Geometrical parameters in $f(R)$ gravity}
\label{section4} The cosmological observations are mainly dependent
on background spatial curvature and four dimensional space-time
curvature of the Universe. In this section we introduce the
geometrical parameters in modified gravity to use it in
observational tests of model.

\subsection{comoving distance}
The radial comoving distance is one of the basic parameters in
cosmology. For an object with a redshift of $z$, using the null
geodesics in the FRW metric, the comoving distance in terms of $X$
is obtained by
\begin{eqnarray} r(z;n,X_0) &=& c\int_0^z\,
{dz' \over H(z')},\\\nonumber &=&
\frac{cH_0^{-1}}{3^{4/3}(\Omega_m)^{1/3}}\int_{X_{p}}^{X}\frac{F'-XF''}{(2F-XF^{'})^{\frac{2}{3}}}\frac{dX}{{\cal
H }(X)},\label{comoving}\nonumber\\
 \end{eqnarray}
where the dimensionless parameter $X$ relates to the redshift from
equation (\ref{apala}) as:
\begin{equation}
z = [\frac{1}{3\Omega_m}(X^n -
X_0^n)^{\frac{1}{n}-1}(X^n-2X_0^n)]^{1/3} - 1
\end{equation}
Knowing the parameters of the action $n$ and $X_0$, we can calculate
the Hubble parameter at a give $X$ by Eq. (\ref{hcal}):substituting
it in (\ref{comoving}) we obtain a comoving distance by numerical
integration. Figures (\ref{fcom1}) and (\ref{fcom2}) show comoving
distance as a function of redshift in the unit of $cH_0^{-1}$ for
various values of parameters of the model. In Fig. (\ref{fcom1}) we
fix $n=1$ which is equivalent to the $\Lambda$CDM universe and let
$X_0$ vary. It seems that $X_0$ plays the role of effective
cosmological constant. Increasing this term makes a larger comoving
distance for a given redshift. In Fig. (\ref{fcom2}) we keep $X_0
=4.2$ and let $n$ change. Increasing the exponent results in a
smaller comoving distance for a given redshift.

\begin{figure}[t]
\epsfxsize=9.5truecm
\begin{center}
\epsfbox{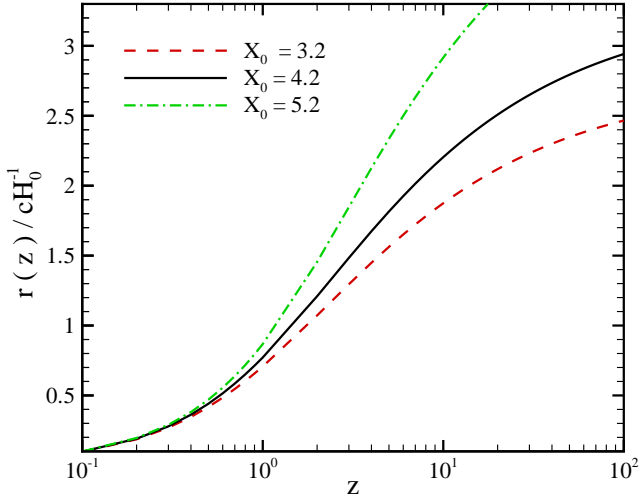}
 \narrowtext \caption{The dependence of comoving
distance as a function of redshift for the case of $n=1$ and various
$X_0$. Increasing $X_0$ makes larger comoving distance for a given
redshift.} \label{fcom1}
\end{center}
\end{figure}
\begin{figure}[t]
\epsfxsize=9.5truecm
\begin{center}
\epsfbox{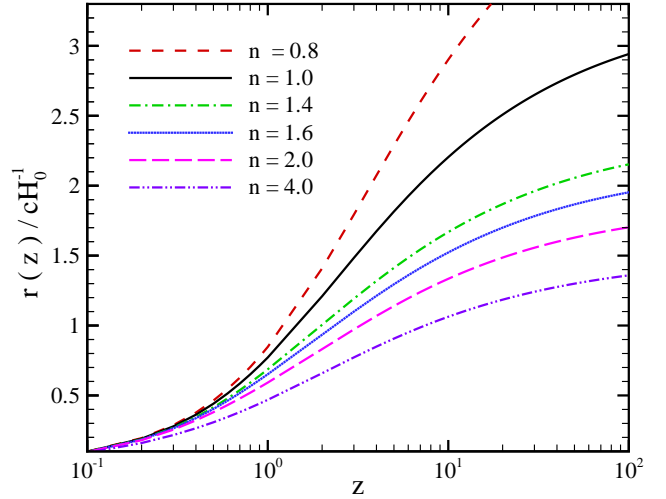}
 \narrowtext \caption{The dependence of comoving
distance as a function of redshift for the case of $X_0=4.2$ and
various $n$. In increasing the exponent results in smaller comoving
distance for a given redshift.} \label{fcom2}
\end{center}
\end{figure}



\subsection{Angular diameter distance and Alcock-Paczynski test}
The apparent angular size of an object located at the cosmological
distance is another important parameter that can be affected by the
cosmological model. An object at the redshift of $z$ and the
perpendicular size of $D_\bot$ is seen by the angular size of
\begin{equation}
\Delta\theta=\frac{D_\bot}{d_A},
 \label{as}
\end{equation}
where $d_A=r(z;n,X_0)/(1+z)$ is the angular diameter distance. Now
imagine this structure has the size of $D_\|$ along our line of
sight. Then the light arriving at us from the back and front of this
structure will not have the same redshift. The difference in the
redshifts of the two sides of the structure can be obtained by the
delay in received light to the observer with $\Delta t(z) = D_\|/c$.
Writing $\Delta t$ in terms of $\Delta a$ as $\Delta t = H^{-1}(z)
\Delta a/a $ we again change $\Delta a$ in terms of $\Delta z$ as
$\Delta z/(1+z) = -\Delta a/a$. The result is writing the width of
the structure in redshift space along our line of sight in terms of
physical size as
\begin{equation}
\Delta z = \frac{1}{c} D_\| H(z) (1+z).
\end{equation}
Now the width of the structure in the redshift space to the apparent
angular size of structure obtain as
\begin{equation}
\frac{\Delta z}{\Delta\theta} = \frac{(1+z) H(z)
d_A}{c}(\frac{D_\|}{D_\bot}). \label{ap}
\end{equation}
For the spherical structures for instance taking into account the
neutral hydrogen clouds at $z<6$ with the spherical symmetric shape,
Eq. (\ref{ap}) is written as:
\begin{equation}
{\Delta z\over \Delta \theta} = \frac{H(z;n,X_0)r(z;n,X_0)}{c}.
\label{alpa}
\end{equation}
This relation is the so-called Alcock-Paczynski test. The advantage
of the Alcock-Paczynski test is that this relation is independent of
the Hubble parameter at the present time and of the existence of the
dust in the intergalactic medium. In this method, instead of using a
standard candle, we will use a standard ruler such as the baryonic
acoustic oscillation.

Figures (\ref{fap1}) and (\ref{fap2}) show a dependence of $\Delta
z/\Delta\theta$ as a function of redshift normalized to the
corresponding value in a $\Lambda$CDM universe. In Fig.(\ref{fap1})
the relative size of the structure in redshift space to the observed
angular size is compared to that in a  $\Lambda$CDM universe for a
fixed value of $n=1$.
In Fig. (\ref{fap2}) we fixed $X_0=4.2$ and change the exponent $n$.
\begin{figure}[t]
\epsfxsize=9.5truecm
\begin{center}
\epsfbox{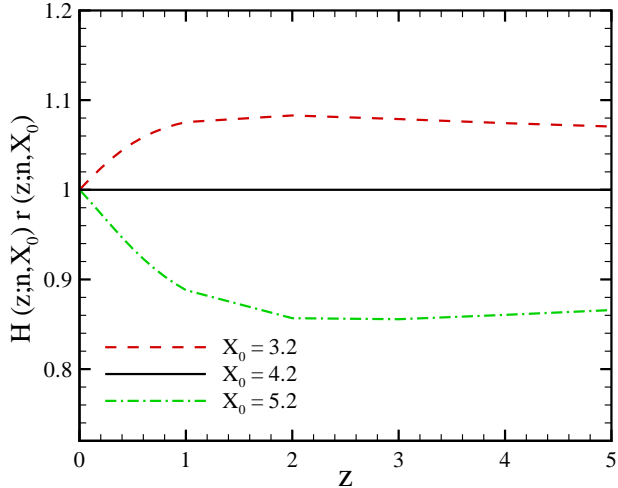}
 \narrowtext \caption{The dependence of $\Delta z/\Delta\theta$ as a
 function of redshift for the case of $n=1$ and various $X_0$.
Increasing $X_0$ causes increasing the apparent size of cosmological
objects.} \label{fap1}
\end{center}
\end{figure}
\begin{figure}[t]
\epsfxsize=9.5truecm
\begin{center}
\epsfbox{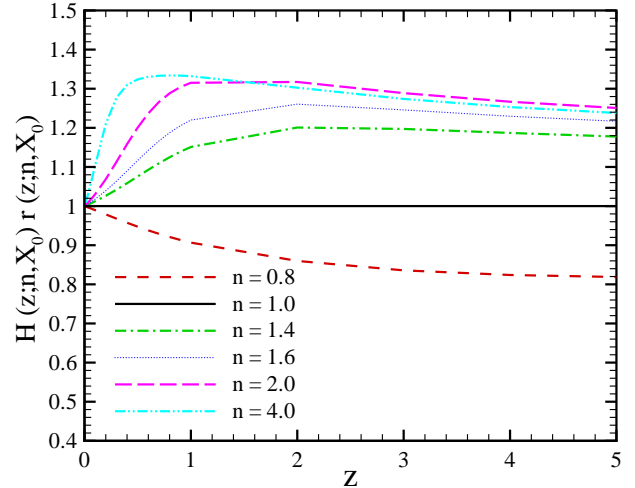}
 \narrowtext \caption{The dependence of $\Delta z/\Delta\theta$
 as a function of redshift for the case of $X_0=4.2$ and various $n$.
Increasing the exponent results in decreasing the apparent
cosmological size of objects.} \label{fap2}
\end{center}
\end{figure}

\subsection{Comoving Volume Element}
The comoving volume element is another geometrical parameter that is
used in number-count tests such as lensed quasars, galaxies, or
clusters of galaxies. The comoving volume element in terms of
comoving distance and Hubble parameter is given by
\begin{equation}
f(z;n,X_0) \equiv {dV\over dz d\Omega} =
\frac{r^2(z;n,X_0)}{H(z;n,X_0)}.
\end{equation}
Figures (\ref{fco1}) and (\ref{fco2}) show the dependence of
comoving volume element as a function of redshift. Figure
(\ref{fco1}) represents the dependence of comoving volume for a
fixed $n=1$ and various $X_0$. Increasing $X_0$ causes larger
comoving volume element. In Fig. (\ref{fco2}) we plot the volume for
fixed $X_0= 4.2$ changing $n$. Increasing the exponent index makes
the comoving volume element smaller.
\begin{figure}[t]
\epsfxsize=9.5truecm
\begin{center}
\epsfbox{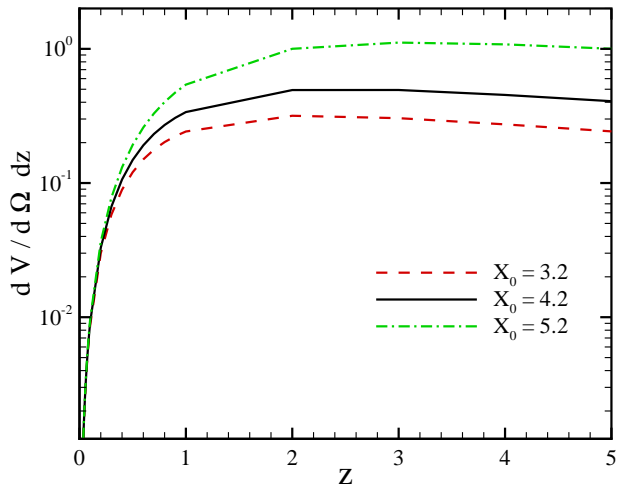}
 \narrowtext \caption{The dependence of comoving element as a
function of redshift for the case of $n=1$ and various $X_0$.
Increasing $X_0$ increases the comoving volume element.}
\label{fco1}
\end{center}
\end{figure}
\begin{figure}[t]
\epsfxsize=9.5truecm
\begin{center}
\epsfbox{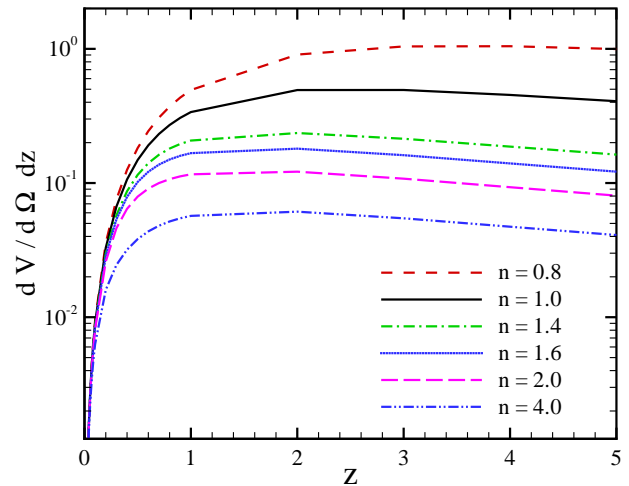}
 \narrowtext \caption{The dependence of comoving element
 as a function of redshift for the case of $X_0=4.2$ and various $n$.
Increasing the exponent causes decreasing the comoving volume
element.} \label{fco2}
\end{center}
\end{figure}

\section{Observational Constraints: Background Evolution}
\label{section5} In this section we compare the observed data with
that from the dynamics of the background from the model. We use
Supernova Type Ia data, CMB-shift parameter, baryonic acoustic
oscillation (BAO) and the gas mass fraction of cluster of galaxies
to constrain the parameters of the model.

\begin{table}
\begin{center}
\caption{\label{tab1} Different priors on the parameter space, used
in the likelihood analysis.}
\medskip
\begin{tabular}{|c|c|c|}
  {\rm Parameter}& Prior & \\  \hline
  $K$ & $0.00$ & {\rm Fixed}\\\hline
 $\Omega_bh^2$&$0.020\pm0.005$&{\rm Top hat (BBN)\cite{bbn}}\\\hline
 $h$&$-$&{\rm Free \cite{hst,zang}}\\\hline
 $w$&$0$&{\rm Fixed}\\
 \end{tabular}
\end{center}
\end{table}

\subsection{Supernova Type Ia}
The Supernova Type Ia experiments provided the main evidence for the
existence of dark energy. Since 1995 two teams of High-Z Supernova
Search and the Supernova Cosmology Project have discovered several
type Ia supernovas at the high redshifts \cite{perl99,sch98}. They
showed that to interpret the faintness of high redshift supernovas
in a flat universe one has to consider an accelerating universe at
the present time.


In this work we take two sets of SNIa data. The first one is the
gold sample which has a 157 supernova {\cite{Riess04} and the second
set is a combined data set of a  192 supernova {\cite{Riess07}}. The
distance modulus for supernovas is calculated by
\begin{eqnarray}
\mu\equiv m-M&=&5\log{D_{L}(z;X_{0},n)}\nonumber\\&&\quad
+5\log{\left(\frac{c/H_0}{1\quad Mpc}\right)}+25, \label{eq:mMr}
\end{eqnarray}
where
\begin{eqnarray}
\label{luminosity} D_L (z;X_{0},n) &=&(1+z)H_0 \int_0^z\, {dz'\over
H(z')},
\end{eqnarray}
and $D_L$ can be written in terms of new parameter $X$,which
appeared in the redefinition of the modified gravity action as
\begin{equation}
D_{L}=\frac{1}{3}\frac{(2F-XF')^{\frac{1}{3}}}{(3\Omega_m)^{\frac{2}{3}}}\int_{X_{p}}^{X}\frac{F'-XF''}{(2F-XF^{'})^{\frac{2}{3}}}\frac{dX}{{\cal
H }(X)}.
\end{equation}
For simplicity in calculation, we define
\begin{eqnarray}
\label{m1} \bar{M} &=& 5\log{\left(\frac{c/H_0}{1\quad
Mpc}\right)}+25,
\end{eqnarray}
which is a function of the  Hubble constant at the present time. We
write the distance modulus as
\begin{eqnarray}
\mu&=&5\log{D_{L}(z;X_{0},n)}+\bar{M}.
\end{eqnarray}

In the next step we use $\chi^2$ fitting to constrain the parameters
of the  model.
\begin{eqnarray}\label{chi_sn}
\chi^2(\bar{M},X_{0},n)&=&\sum_{i}\frac{[\mu_{obs}(z_i)-\mu_{th}(z_i;\bar{M},X_0,n)]^2}{\sigma_i^2},\nonumber\\
\end{eqnarray}
where $\sigma_i$ is the uncertainty in the distance modulus. To
constrain the parameters of the  model, we use the likelihood
statistical analysis
\begin{eqnarray}
{\cal L}(\bar{M},X_{0})={\mathcal{N}}e^{-\chi^2(\bar{M},X_{0})/2},
\end{eqnarray}
where ${\mathcal{N}}$ is a normalization factor. The parameter
$\bar{M}$ is a nuisance parameter and should be marginalized
(integrated out) leading to a new $\bar{\chi}^{2}$ defined as:
\begin{eqnarray}\label{mar2}
\bar{\chi}^2=-2\ln\int_{-\infty}^{+\infty}e^{-\chi^2/2}d\bar{M}.
\end{eqnarray}
 Using Eqs. (\ref{chi_sn}) and (\ref{mar2}), we find
\begin{eqnarray}\label{mar3}
\bar{\chi}^2(X_{0})&=&\chi^2(\bar{M}=0,X_{0})-\frac{B(X_{0})^2}{C}+\ln(C/2\pi),
\end{eqnarray}
where
\begin{eqnarray}\label{mar4}
B(X_{0})=\sum_{i}\frac{[\mu_{obs}(z_i)-\mu_{th}(z_i;X_{0},\bar{M}=0)]}{\sigma_i^2},
\end{eqnarray}
and
\begin{eqnarray}\label{mar5}
C=\sum_{i}\frac{1}{\sigma_i^2}.
\end{eqnarray}
Equivalent to marginalization is the minimization of $\chi^2$ with
respect to $\bar{M}$. One can show that $\bar\chi^2$ can be expanded
in terms of $\bar{M}$ :
\begin{eqnarray}\label{mar6}
\chi^2_{\rm
SNIa}(X_{0})=\chi^2(\bar{M}=0,X_{0})-2\bar{M}B+\bar{M}^2C,
\end{eqnarray}
which has a minimum value for $\bar{M}=B/C$ and results in:
\begin{eqnarray}\label{mar7}
\chi^2_{\rm
SNIa}(X_{0})=\chi^2(\bar{M}=0,X_{0})-\frac{B(X_{0})^2}{C}.
\end{eqnarray}
Using equation (\ref{mar7}) we can find the best fit values of model
parameters, minimizing $\chi^2_{\rm SNIa}(X_{0})$.
\begin{figure}[t]
\epsfxsize=9.5truecm\epsfbox{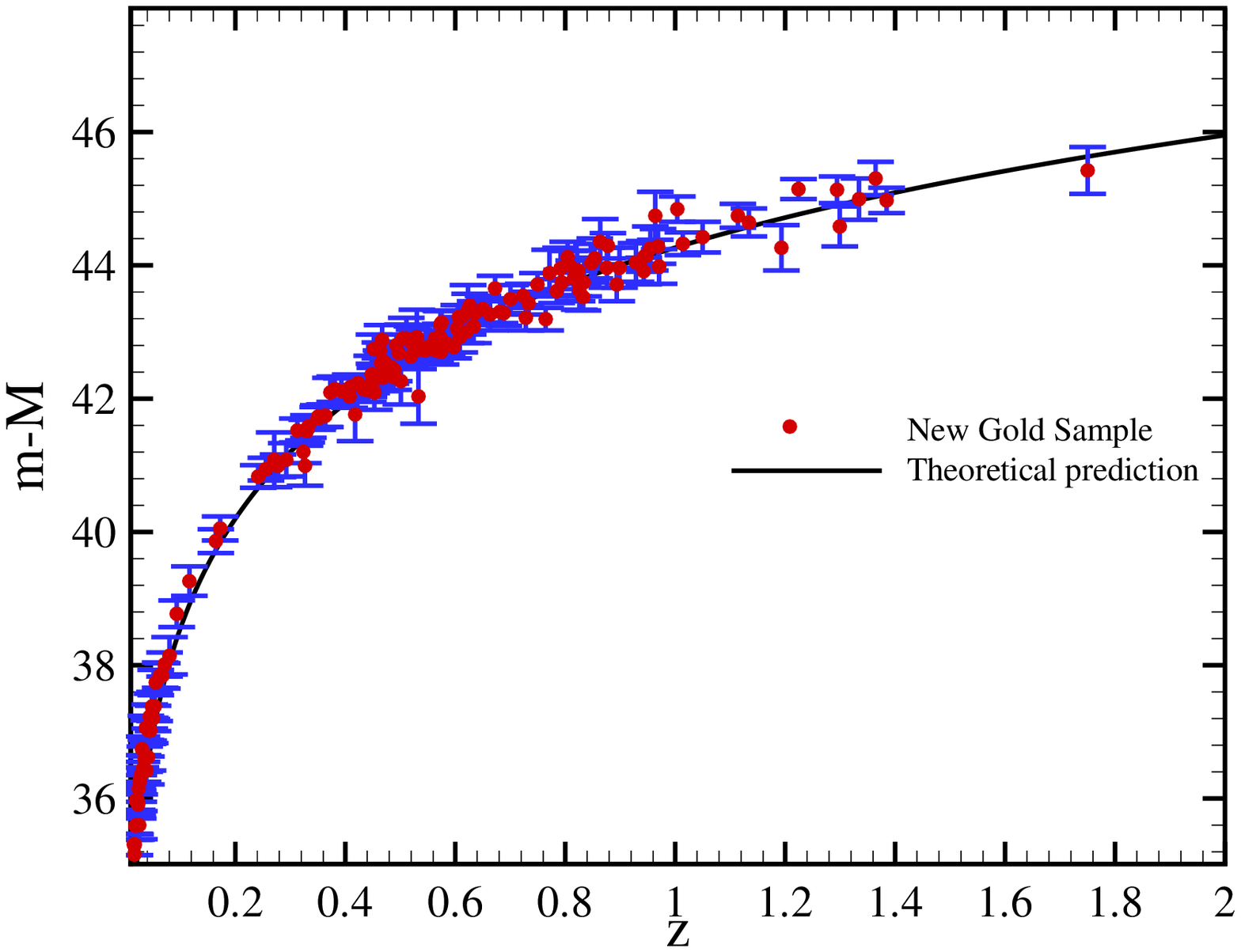} \narrowtext \caption{The
best fit of distance modulus as a function of redshift to the
Supernova Type Ia new gold sample.} \label{sngold}
\end{figure}
\begin{figure}[t]
\epsfxsize=9.5truecm\epsfbox{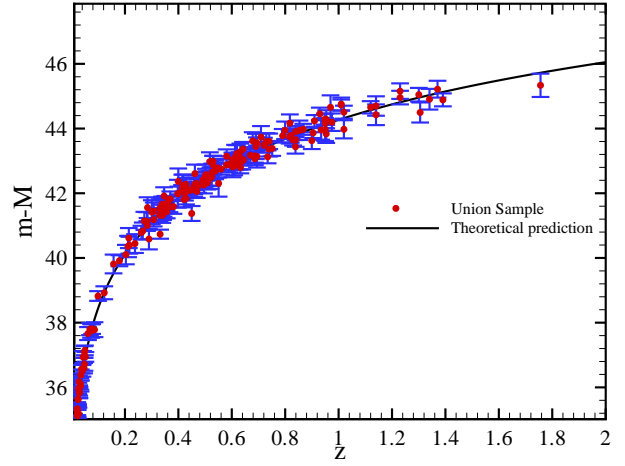} \narrowtext \caption{The best
fit of distance modulus as a function of redshift to the Supernova
Type Ia Union sample.} \label{snWR}
\end{figure}
Figures (\ref{sngold}) and (\ref{snWR}) represent the best fit to
the Supernova Type Ia new gold sample and union sample respectively.
The best fit values for the free parameter of the model for two
cases are $ n = 2.01 ^{+0.72}_{-0.67}, X_0 = 6.45^{+1.13}_{-1.51}$
and $\Omega_m = 0.67^{+0.64}_{-0.64}$ for the new gold sample and $
n = 1.63 ^{+0.76}_{-0.92}, X_0 = 6.09^{+1.32}_{-2.86}$ and $\Omega_m
= 0.47^{+0.69}_{-0.47}$ for mixed Gold-SNLS data. Figures
(\ref{lik-sn1}) and (\ref{lik-sn2}) represent the likelihood
functions in terms of $n$ and $X_0$.

\begin{figure}[t]
\epsfxsize=9.5truecm\epsfbox{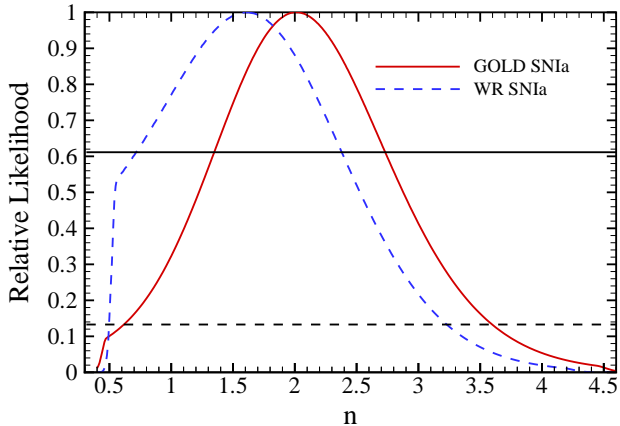} \narrowtext
\caption{Marginalized likelihood functions of $f(R)$ modified
gravity model free parameter, $n$. The solid and dash lines
correspond to the likelihood function of fitting the model with SNIa
data new gold sample and union data set, respectively. The
intersections of the curves with the horizontal solid and dashed
lines give the bounds with $1\sigma$ and $2\sigma$ level of
confidence, respectively.} \label{lik-sn1}
\end{figure}

\begin{figure}[t]
\epsfxsize=9.5truecm\epsfbox{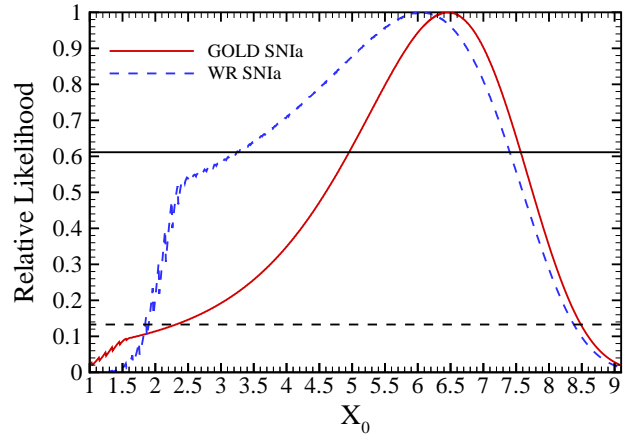} \narrowtext
\caption{Marginalized likelihood functions of $f(R)$ modified
gravity model free parameter, $X_0$. The solid and dash lines
correspond to the likelihood function of fitting the model with SNIa
data new gold sample and union data set, respectively. The
intersections of the curves with the horizontal solid and dashed
lines give the bounds with $1\sigma$ and $2\sigma$ level of
confidence, respectively.} \label{lik-sn2}
\end{figure}

\subsection{CMBR Shift parameter}
Another dynamical parameter that is used in recent cosmological
tests is the CMB shift parameter. Before the last scattering epoch,
the baryons and photons were tightly coupled through the
electromagnetic interaction. This coupled fluid was under the
influence of two major forces of (a) the gravitational pull of
matter and (b) the out leading pressure of photons. The finger print
of this competition leads to the familiar spectrum of peaks and
troughs on the CMB map. Here the main peak is the so-called acoustic
peak. The odd peaks of the CMB anisotropy spectrum correspond to the
maximum compression of the fluid, the even ones to the rarefaction
\cite{Hu97}.

In an idealized model of the fluid, there is an analytic relation
for the location of the m-th peak: $l_m \approx ml_A$
\cite{Hu95,hu00}, where $l_A$ is the acoustic scale which may be
calculated analytically and depends on both pre- and
post-recombination physics as well as the geometry of the universe.
The acoustic scale corresponds to the Jeans length of photon-baryon
structures at the last scattering surface some $\sim 379$ Kyr after
the big bang \cite{spe03}. The apparent angular size of the acoustic
peak can be obtained by dividing the comoving size of the sound
horizon at the decoupling epoch $r_s(z_{dec})$ by the comoving
distance of observer to the last scattering surface $r(z_{dec})$:
\begin{equation}
\theta_A =\frac{\pi}{l_A}\equiv {{r_s(z_{dec})}\over r(z_{dec}) }.
\label{eq:theta_s}
\end{equation}
The nominator of Eq. (\ref{eq:theta_s}) corresponds to the distance
that the perturbation of pressure can travel from the big bang to up
to the last scattering surface, which is defined as the integral
below:
\begin{equation}
 r_{s}(z_{dec})= \int_{z_{dec}}^ {\infty} {v_s(z')dz'\over H(z')/H_0} \label{sh}
 \end{equation}
where $v_s(z)^{-2}=3 + 9/4\times\rho_b(z)/\rho_{rad}(z)$ is the
sound velocity in the unit of speed of light from the big bang up to
the last scattering surface \cite{Hu95,doran} $z_{dec}$ is the
redshift of the last scattering surface.

Changing the parameters of the model can change the size of the
apparent acoustic peak and subsequently the position of $l_A\equiv
\pi/\theta_A$ in the power spectrum of temperature fluctuations on
CMB. The simple relation $l_m\approx ml_A$ however does not hold
very well for the first peak although it is better for higher peaks
[2]. Driving effects from the decay of the gravitational potential
as well as contributions from the Doppler shift of the oscillating
fluid introduce a shift in the spectrum. A good parameterizations
for the location of the peaks and troughs is given by
\begin{equation}\label{phase shift}
l_m=l_A(m-\phi_m)
\end{equation}
where $\phi_m$ is a phase shift determined predominantly by
prerecombination physics, and is independent of the geometry of the
Universe. Instead of the peak locations of the power spectrum of
CMB, one can use another model-independent parameter, which is the
so-called shift parameter ${\cal R}$, as
\begin{equation}
{\cal R}= \frac{\omega_m^{1/2}}{\omega_k^{1/2}}sinn_k(\omega_k r)
\end{equation}
where $sinn_k(x) = \sin(x), x, \sinh(x)$ for $k=-1,0,1$. For the
case of a flat universe, which is our concern, the shift parameter
reduces to the simpler formula of
\begin{equation}
{\cal{R}}=\sqrt{\Omega_{m}H_{0}^{2}}\int_{0}^{z_{dec}}\frac{dz}{H(z)}.
\end{equation}
Now we change the variable from the redshift to $X$ and rewrite the
above expression in terms of dimensionless parameter $X$ and take
the integral from the value of $X$ at the present time as the lower
limit of the integral and the value of $X$ at the decoupling time as
the upper limit:
\begin{equation}
{\cal{R}}=\frac{\Omega_{m}^{\frac{1}{6}}}{3^{\frac{4}{3}}}\int_{X_{p}}^{X_{dec}}
    \frac{F'-XF''}{(2F-XF^{'})^{\frac{2}{3}}}\frac{dX}{{\cal{H}}(X)}.
\end{equation}
The observed result of the CMB experiment is
${\cal{R}}=1.716\pm0.062$ \cite{spe03}. It is worthwhile to mention
that the dimensionless parameter ${\cal{R}}$ is independent of the
Hubble constant. We compare the observed shift parameter with that
of the model using the likelihood analyzing, minimizing $\chi^2$
defined as
\begin{equation}\label{chi_cmb}
\chi_{CMB}^2=\frac{\left[{\cal R}_{{\rm obs}}-{\cal R}_{{\rm
the}}\right]^2}{\sigma_{\rm CMB}^2}.
\end{equation}
\subsection{Baryon Acoustic Oscillations}
Another geometrical cosmological probe which determines the
distance-redshift relation is BAO. The physics governing the
production of BAO is well understood. Acoustic peaks occurred
because cosmological perturbations excite sound waves in initial
relativistic plasma in the early epoch of the Universe. Dark matter
perturbations grows in place while the baryonic matter perturbations
were carried out in an expanding spherical wave because of their
interaction with photons. At the recombination epoch, when the
photons started to decouple from the baryonic matter, the shell of
the baryonic matter perturbation sphere was nearly 150 Mpc in the
comoving frame. From the linear structure formation theories, this
scale should not be changed until the present time. The structure
formation theory predicts that this 150Mpc imprint of baryonic
matter remains in the correlation function of the density contrast
and can be seen in the large-scale surveys.

By knowing the size of acoustic oscillation, one can measure the
angular distance to this structure. The large-scale correlation
function measured from 46748 luminous red galaxies spectroscopic
sample of SDSS include a clear peak at 100 Mpc$h^{-1}$ \cite{Eis05}.
The corresponding comoving scale of the sound horizon shell is about
150Mpc in radius. A dimensionless and $H_{0}$ independent parameter
for constraining the cosmological models has been proposed in
literatures \cite{Eis05} as follows:
\begin{equation} \label{lss1}
{\cal A} = \sqrt{\Omega_m}\left[\frac{H_0D_L^2(z_{\rm
sdss};X_0)}{H(z_{\rm sdss};X_0)z_{\rm sdss}^2(1+z_{\rm
sdss})^2}\right]^{1/3}.
\end{equation}
or in simpler form
\begin{equation}
{\cal
A}=\sqrt{\Omega_{m}}{\cal{H}}(X)^{-\frac{1}{3}}{\left[\frac{1}{z_{\rm
sdss}}\int_{0}^{z_{\rm
sdss}}\frac{dz}{{\cal{H}}(X)}\right]}^{\frac{2}{3}}.
\end{equation}
We rewrite the above dimensionless quantity in terms of modified
gravity model parameters as
\begin{eqnarray}
{\cal A}&=&\sqrt{\Omega_{m}}{\cal{H}}(X)^{-\frac{1}{3}}\nonumber\\
&&\times \left[\frac{(3\Omega_m)^{-\frac{1}{3}}}{3z_{\rm
sdss}}\int_{X_{p}}^{X_{\rm
sdss}}\frac{(F'-XF'')dX}{{\cal{H}}(X)(2F-XF')^{2/3}}\right]^{\frac{2}{3}}
\end{eqnarray}
Now we can put a constraint on the $f(R)$ modified gravity model
using the value of ${\cal{A}}=0.469\pm0.017$ from luminous red
galaxies observation at $z_{SDSS}=0.35$ \cite{Eis05}. It is
worthwhile to mention that the procedure above presented in
literature is well proposed for dark energy models in which the
$\Omega_{m}$ has the same definition in standard cosmology. In
contrast to the dark energy models in gravity theories in the
Palatini formalism, $\Omega_{m}$ is a conventional dimensionless
parameter and does not have the same role as in dark Energy models,
considering the well-known fact that $\Omega_{m}=1$ does not
correspond to a flat universe as in standard FRW equations.
Consequently in order to not include the weak model dependence of
the dimensionless parameter, ${\cal{A}}$, we will use another
similar approach, proposed by Percival et al. \cite{Perc08}. This
method constrain general cosmological models by using BAO distance
measurement from galaxy samples covering different redshift ranges.
Measuring the distance redshift relation at two redshifts of $z=0.2$
and  $z=0.35$ for clustering of SDSS luminous red galaxies enables
us to define a new dimensionless parameter as
\begin{equation}
{\cal{B}}=\frac{D_{V}(z=0.35)}{D_{V}(z=0.20)}
\end{equation}
where $D_{V}$ is given by
\begin{equation}
D_{V}=\left[{\frac{(1+z)^{2}d_{A}^{2}cz}{H(z)}}\right]^{1/3},
\end{equation}
and $d_{A}$ is the angular diameter distance. The observational
values for two different redshifts are reported in \cite{Perc08}
with 1$\sigma$ error:
\begin{eqnarray}
\frac{r_{s}}{D_{V}(z=0.20)}=0.1980\pm0.0058\\
\frac{r_{s}}{D_{V}(z=0.35)}=0.1094\pm0.0033\\
\end{eqnarray}
where $r_{s}$ is the comoving sound horizon scale at the
recombination epoch.Considering that BAO measurements have the same
measured scale at all redshifts then we have a numerical value for
${\cal{B}}$ as
\begin{equation}
{\cal{B}}=\frac{D_{V}(z=0.35)}{D_{V}(z=0.2)}=1.812\pm0.060.
\end{equation}
Now we convert the comoving angular diameter distance to luminosity
distance $D_{L}$ and calculate $D_{L}$ in terms of the dimensionless
Hubble parameter in modified gravity
\begin{equation}
{\cal{B}}=\left[{\frac{{\cal{H}}(z=0.20){D_{L}}^{2}(z=0.35)0.35(1+0.2)^{2}}{{\cal{H}}(z=0.35){D_{L}}^{2}(z=0.20)0.2(1+0.35)^2}}\right]^{1/3}.
\end{equation}
We use $\chi^2$ as one more fitting parameter with the observed
value of ${\cal{B}}=1.812\pm0.060$. This observation permits us to
add one more term to $\chi^2$ from that of SNIa and CMB-shift
parameter by minimizing
\begin{equation}
{\chi}^2_{BAO}=\frac{({\cal{B}}_{obs}-{\cal{B}}_{th})^2}{\sigma^2_{BAO}}.
\end{equation}
This is the third geometrical parameter we will use to constrain the
model.
\subsection{Gas mass fraction of cluster of galaxies}
Measurement of the ratio of X-ray emitting gas to the total mass in
galaxy clusters ($f_{gas}$) also is an indication of the
acceleration of the Universe. This method can be used as another
cosmological test to constraint the parameters of the model. Galaxy
clusters are the largest objects in the Universe; the gas fraction
in them is presumed to be constant and nearly equal to the baryon
fraction in the Universe. Sasaki (1996) and Pen (1997) described how
measurements of the apparent dependence of the baryonic mass
fraction could also, in principle, be used to constrain the geometry
and matter content of a universe \cite{Sasaki96,pen97}. The
geometrical constraint arises from the dependance of the measured
baryonic mass fraction value on the assumed angular diameter
distance to the clusters \cite{Allen04}. The baryonic mass content
of galaxy clusters is dominated by the X-ray emitting intercluster
gas, the mass of which exceeds the mass of optically luminous
material by a factor of $6$ \cite{white93,Fuk98}. Let us define
$f_{gas}$ as
\begin{equation}\label{fgas}
f_{gas}=\frac{M_{gas}}{M_{tot}}
\end{equation}
In the second step we want to replace the mass of gas by the
baryonic mass considering that:
\begin{equation}
M_{b}=(1+\beta)M_{gas},
\end{equation}
where from the observations we know $\beta=0.19h^{1/2}$
\cite{white93}. On the other hand we assume that we are observing
rich cluster of galaxies where the fraction of baryonic mass to the
total mass has the same fraction as in the universe with a bias
factor $b$. We substitute this assumption in Eq. (\ref{fgas}) to
achieve
\begin{equation}
f_{gas}=\frac{b}{1+\beta}\frac{\Omega_{b}}{\Omega_{m}}.
\label{fgas2}
\end{equation}
Using the distribution of gas and matter in cluster, Sasaki (1996)
showed that the fraction of gas depends on angular distance with
$f_{gas}\propto D_{A}^{3/2}$ \cite{Sasaki96}. On the other hand the
fraction of gas obtained from the observation depends on the model
we are assuming for the dynamics of the universe. It is assumed that
$f_{gas}$ should in reality be independent of the redshift. To
determine the constraints on the proposed modified gravity action,
we fit the $f_{gas}$ data with a model that accounts for the
expected apparent variation in $f_{gas}(z)$ as the underlying
cosmology is varied. We choose both SCDM (a flat universe with
$\Omega_m = 1$ and $h=0.5$) and $\Lambda$CDM as reference cosmology
models. The ratio of gas fraction for a given model to the reference
model is: $f_{gas}^{(ref)}/f_{gas}^{(mod)} =
\left[D_A^{(ref)}/D_A^{(mod)}\right]^{3/2}$. On the other hand using
Eq. (\ref{fgas2}) for a given model, the gas fraction for a
reference model is obtained by:
\begin{equation}
f^{(ref)}_{gas}=\frac{b\Omega_{b}}{(1+0.19\sqrt{h})\Omega_{m}}[\frac{D^{ref}_{A}(z)}{D^{mod}_{A}(z)}]^{3/2},
\end{equation}
where superscrips $(ref)$ correspond once to SCDM and then to the
$\Lambda$CDM model \cite{all04}. We use $\chi^2$ to compare gas
fractions of observational and theoretical models as follows:
\begin{equation}
\chi_{gas}=\frac{(f^{obs}_{gas}-f^{the}_{gas})^{2}}{\sigma^2_{gas}}.
\end{equation}
This is the fourth geometrical constraint.

\subsection{Combined analysis: SNIa$+$CMB$+$BAO$+$GAS-FRACTION} \label{cmb}
\begin{figure}
\epsfxsize=9.5truecm\epsfbox{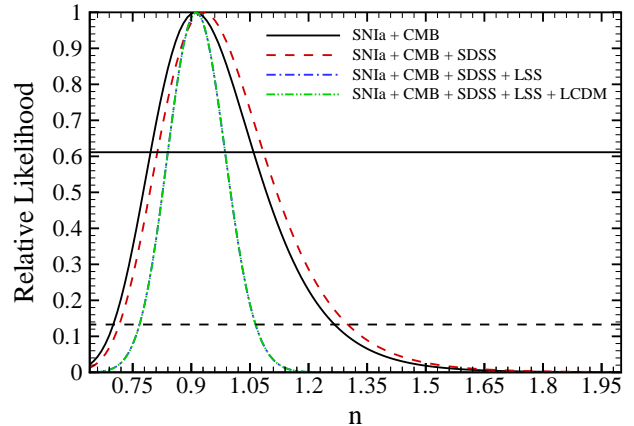} \narrowtext
\caption{Marginalized likelihood functions of $f(R)$ modified
gravity model free parameter, $n$. The solid line corresponds to the
joint analysis of SNIa data (new gold sample) and CMB, the dashed
line shows the joint analysis of SNIa$+$CMB$+$SDSS data,the dash-dot
line corresponds to SNIa$+$CMB$+$SDSS$+$LSS and dash-dot-dot line
indicates SNIa$+$CMB$+$SDSS$+$LSS$+$LCDM. The intersections of the
curves with the horizontal solid and dashed lines give the bounds
with $1\sigma$ and $2\sigma$ level of confidence respectively. The
results for two former analyses are very similar.} \label{geolike1}
\end{figure}

\begin{figure}[t]
\epsfxsize=9.5truecm\epsfbox{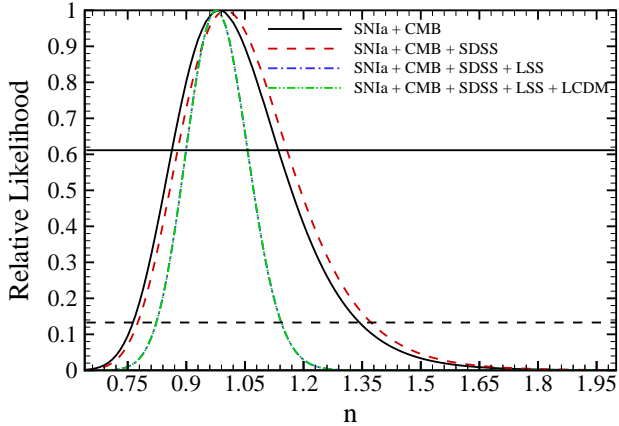} \narrowtext
\caption{Marginalized likelihood functions of $f(R)$ modified
gravity model free parameter, $n$. The solid line corresponds to the
joint analysis of SNIa data and CMB, the dashed line shows the joint
analysis of SNIa$+$CMB$+$SDSS data, dash-dot line corresponds to
SNIa$+$CMB$+$SDSS$+$LSS and the dash-dot-dot line indicates
SNIa$+$CMB$+$SDSS$+$LSS$+$LCDM. The intersections of the curves with
the horizontal solid and dashed lines give the bounds with $1\sigma$
and $2\sigma$ level of confidence, respectively. The results for two
former analysis are very similar.} \label{geolike2}
\end{figure}

In this section we combine SNIa data (from SNIa new Gold sample and
mixed SNLS), CMB shift parameter from the WMAP, recently observed
baryonic peak from the SDSS and 2dF and the gas mass fraction in
cluster of galaxies to constrain the parameter of the modified
gravity model by minimizing the combined $\chi^2 = \chi^2_{\rm
{SNIa}}+\chi^2_{{\rm CMB}}+\chi^2_{{\rm BAO}}+\chi^2_{\rm{gas}}$.

The best values of the parameters of the model from the fitting with
data, including SNIa new sample are $n = 0.91^{+0.08}_{-0.07}$, $X_0
= 3.67^{+0.44}_{-0.42}$ and $\Omega_m = 0.29^{+0.10}_{-0.09}$ and
data with including SNIa union sample results in $n =
0.98^{+0.08}_{-0.08}$, $X_0 = 4.39^{+0.38}_{-0.42}$ and $\Omega_m =
0.25^{+0.10}_{-0.010}$. Here we marginalized overall Hubble
parameter in likelihood analysis. Figures (\ref{geolike1}) and
(\ref{geolike2}) show the likelihood function as function of
exponent $n$.
Also Figures (\ref{geolike3}) and (\ref{geolike4}) represent the
likelihood function of $X_0$ in two different supernova data sets.

\section{Constraints by Large-Scale structures:Dynamical parameter  }
\label{section6}
 So far we have only considered the observational results
related to the background evolution. In this section, using the
linear approximation of structure formation, we obtain the growth
index of structures and compare it with the result of observations
by the $2$-degree Field Galaxy Redshift Survey ($2$dFGRS). As we
mentioned before, the evolution of the structures depends on both
the dynamics of the background and the gravity law that governs the
dynamics of particles inside the structure.


Here the evolution of structures in the modified gravity will be
studied through the spherical collapse model. Recently a procedure
has been put forward by Lue, Scoccimarro, and Starkman (2004) which
relies on the assumption that Birkhoff's theorem, holds in a more
general setting of modified gravity theories. This procedure also is
applied in the Palatini formalism of $f(R)$ gravity \cite{tavakkol}.
According to this procedure, it is assumed that the growth of
large-scale structure can be modeled in terms of a uniform sphere of
dust of constant mass. This structure evolves as a FRW universe.
Using Birkhoff's theorem, the space-time at the empty exterior of
this structure is then taken to be a Schwarzschild-like metric. The
components of the exterior metric are then uniquely determined by
smoothly matching the interior and exterior regions. In the Palatini
formalism the metric outside the spherical distribution of matter
depends on the density of matter which may modify the Newtonian
limit of these theories, however here we assume Schwarzschild-like
Newtonian limit.

\begin{figure}[t]
\epsfxsize=9.5truecm\epsfbox{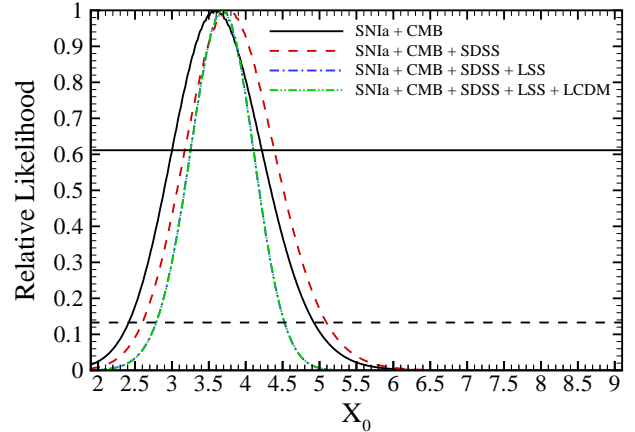} \narrowtext
\caption{Marginalized likelihood functions of $f(R)$ modified
gravity model as a function of $X_0$. The solid line corresponds to
the joint analysis of SNIa data (new Gold sample) and CMB, the
dashed line shows the joint analysis of SNIa$+$CMB$+$SDSS data,
dash-dotted line corresponds to SNIa$+$CMB$+$SDSS$+$LSS and
dash-dot-dot line indicates SNIa$+$CMB$+$SDSS$+$LSS$+$LCDM. The
intersections of the curves with the horizontal solid and dashed
lines give the bounds with $1\sigma$ and $2\sigma$ level of
confidence respectively. The results for two former analysis are
very similar.} \label{geolike3}
\end{figure}

\begin{figure}[t]
\epsfxsize=9.5truecm\epsfbox{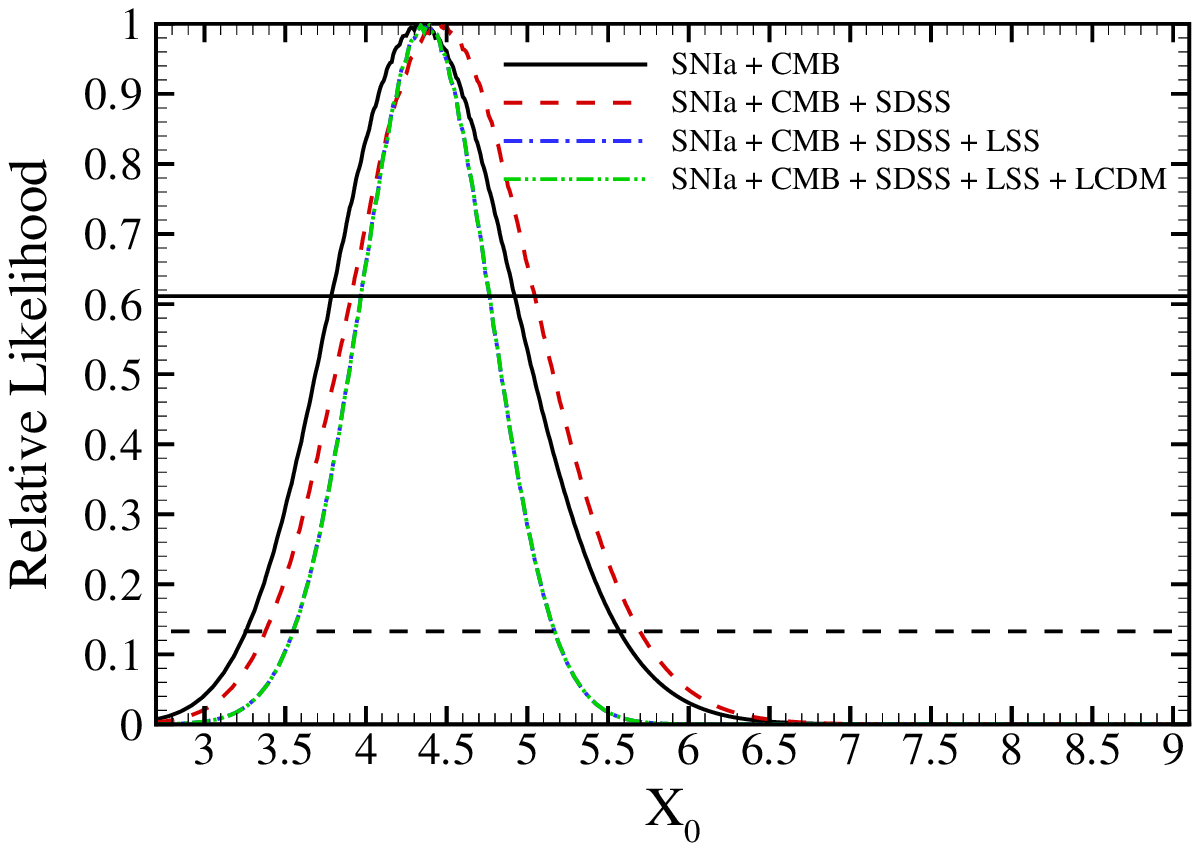} \narrowtext
\caption{Marginalized likelihood functions of $f(R)$ modified
gravity model as a function of $X_0$. The solid line corresponds to
the analysis of SNIa data and CMB, the dashed line shows the joint
analysis of SNIa$+$CMB$+$SDSS data, dash-dotted line corresponds to
SNIa$+$CMB$+$SDSS$+$LSS and dash-dot-dot line indicates
SNIa$+$CMB$+$SDSS$+$LSS$+$LCDM. The intersections of the curves with
the horizontal solid and dashed lines give the bounds with $1\sigma$
and $2\sigma$ level of confidence respectively. The results for two
former analysis are very similar.} \label{geolike4}
\end{figure}

The continuity and Poisson equations for the density contrast
$\delta=\delta \rho/\bar{\rho}$ in the cosmic fluid provides the
evolution of density contrast in the linear approximation (i.e.
$\delta\ll 1 $) as
\begin{equation}
\ddot{\delta}+2\frac{\dot{a}}{a}\dot{\delta} - 4\pi G \rho\
\delta=0, \label{eq2}
\end{equation}
where the dot denotes the time derivative and we assume the size of
structures to be larger than the Jeans length. The effect of a
modified gravity results from the modification to the background
dynamics and we adopt the same Poisson equation for the weak field
regime. In order to use the constraint from large-scale structure,
we rewrite the above equation in terms of $X$. So we have
\begin{eqnarray}
\label{fu2}
\frac{d^2\delta}{da^2}+\frac{d\delta}{da}\left[\frac{3}{a}+\frac{{\cal
H}'(X)}{{\cal
H}(X)}\frac{dX}{da}\right]-\frac{3\Omega_m}{2{\cal{H}}^2(X)a^5}\delta=0
\end{eqnarray}

In the standard linear perturbation theory, the peculiar velocity
field $\bf{v}$ is determined by the density contrast  as
\begin{equation}
 {\bf v} ({\bf x})= H_0 \frac{f}{4\pi} \int \delta ({\bf y})
\frac{{\bf x}-{\bf y}}{\left| {\bf x}-{\bf y} \right|^3} d^3 {\bf
y}, \label{d}
\end{equation} where the growth index $f$ is defined by
\begin{equation} f=\frac{d \ln \delta}{d \ln a},
 \label{eq5}
\end{equation}
and it is proportional to the ratio of the second term of Eq.
(\ref{eq2}) (friction) to the third term (Poisson).

We use the evolution of the density contrast $\delta$ to compute the
growth index of structure $f$, which is an important quantity for
the interpretation of peculiar velocities of galaxies. Replacing the
density contrast with the growth index in Eq.(\ref{d}) results in
the evolution of growth index as
%
%
%
%
\begin{eqnarray}
\label{index1} &&\frac{df}{d\ln a}= \frac{3\Omega_m}{2a{\cal
H}^2(X)}-f^2-f\left[2+\frac{a{\cal H}'(X)}{{\cal
H}(X)}\frac{dX}{da}\right].
\end{eqnarray}
To put constraint on the model using large structure data, we rely
on the observation of $220000$ galaxies with the $2$dFGRS
experiment, which provides a numerical value for the growth index.
By measurements of the two-point correlation function, the $2$dFGRS
team reported the redshift distortion parameter of $\beta = f/b
=0.49\pm0.09$ at $z=0.15$, where $b$ is the bias parameter
describing the difference in the distribution of galaxies and their
masses. Verde et al. (2003) used the bispectrum of $2$dFGRS galaxies
\cite{ver1,ver2} and obtained $b_{verde} =1.04\pm 0.11$ which gave
$f= 0.51\pm0.10$. Now we fit the growth index at $z = 0.15$ derived
from the Eq.(\ref{index1}) with the observed value.
\begin{eqnarray}
\chi^2_{\rm
{LSS}}=\frac{[f_{obs}(z=0.15)-f_{th}(z=0.15;X_0)]^2}{\sigma_{f_{obs}}^2}.
\end{eqnarray}

Finally we do likelihood analysis with considering all the
observations and obtain the 2D distribution of a likelihood function
in terms of $n$ and $X_0$ in Fig.(\ref{2d}).

\subsection{Perturbation Theory} In the previous section we have seen
the effect of modified gravity on the structure formation in the
weak filed regime through the background effect. In this method
changing the dynamics of universe (i.e. scale factor) alters the
formation of the large-scale structures.

In this section we study the relativist structure formation theory
through the perturbation in the homogenous background metric and
energy-momentum tensor. This approach may uncover whether modified
gravity theories driving late-time acceleration predict any testable
features on CMB or large-scale structures in linear or nonlinear
regimes.

Let us consider a flat universe dominated by pressureless cold dark
matter. We identify perturbation in conformally flat FRW space-time
by ten elements as follows:
\begin{eqnarray}
ds^2=a^2(\eta)\{-(1+2\alpha){d\eta}^2-2(\beta_{,i}+b_{i})d\eta{dx^{i}}\nonumber\\
+[g_{ij}^{3}+2(g_{ij}^{(3)}\phi+\gamma_{,ij}+c_{(i,j)})+h_{ij}]dx^{i}dx^{j}
\}
\end{eqnarray}
where $\eta$ is the conformal time. $\alpha$,$\beta$,$\gamma$,and
$\phi$ are the scalar perturbations to the metric. $b_{i}$ and
$c_{i}$ are divergenceless vectors where each one with 2 degrees of
freedom and $h_{ij}$ is a traceless--divergenceless symmetric
$3\times3$ matrix with 2 degree of freedom.

On the other hand perturbation of conservation of energy-momentum
tensor results in the continuity and Euler equations as follows:
\begin{eqnarray}
\dot{\delta}&=&-kv+a\kappa-3H\alpha,\\
\dot{v}&=&-Hv+k\alpha,
\end{eqnarray}
where $\kappa=3a^{-1}(H\alpha-\dot{\phi})-a^{-2}{\Box \chi}$ and
$\chi=a(\beta+\dot{\gamma})$.

Using conformal gauge, the field equation for the density contrast
in the modified gravity framework obtain as \cite{koivi06}:
\begin{equation} \label{density}
\delta^{''}+\xi H\delta^{'}-\zeta(\frac{H^{''}}{H}-2H^{'})\delta=0
\end{equation}
where $'$ is derivative with respect to the conformal time and
$H=\frac{a^{'}}{a}=\dot{a}$. $\xi$ and $\zeta$ are defined as
\begin{eqnarray}
\xi&=&1+\frac{2FF^{''}H-2{F^{'}}^{2}H-2FF^{'}H^{'}}{FH^{2}(2FH+F^{'})},\\
\zeta&=&1+\frac{H^2-H^{'}}{H^{''}-2H^{'}H}(1-\xi)\nonumber\\
&-&\frac{F^{'}H}{3(2FH+F^{'})(H^{''}-2H^{'}H)}k^{2}, \label{zeta}
\end{eqnarray}
where $k$ is the wave number of structures in the universe.
\begin{figure}
\epsfxsize=8.5truecm
\begin{center}
\epsfbox{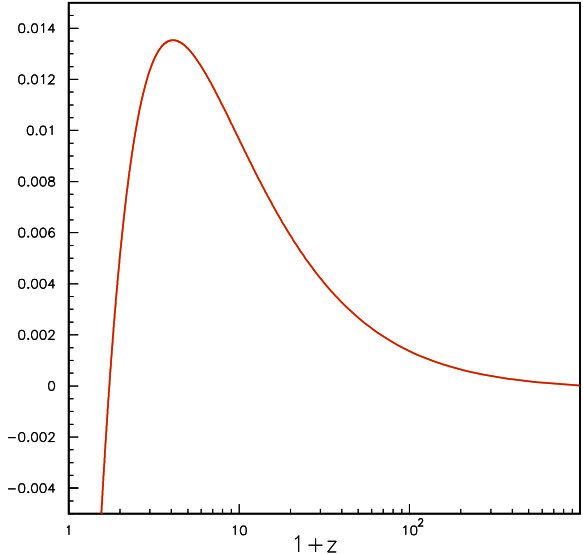}\narrowtext\epsfxsize=8.5truecm
\caption{Difference between the density contrast in $\Lambda$CDM
model and Modified gravity model. For larger redshifts the
difference between these two solutions is negligible.} \label{pert}
\end{center}
\end{figure}
In the case of the Einstein-Hilbert action, $F'=F''=0$ which results
in $\xi =1$ and subsequently $\zeta =1$. The differential equation
governing the evolution of the density contrast in this case reduces
to:
\begin{equation}
\delta^{''}+H\delta^{'}-(\frac{H^{''}}{H}-2H^{'})\delta=0.
\label{perlcdm}
\end{equation}
For comparison of Eqs.(\ref{density})and (\ref{perlcdm}), we obtain
the difference in the density contrast between the $\Lambda$CDM
model and the modified gravity as indicated in Fig.(\ref{pert}) for
a structure with the size of $k = 0.01 Mpc^{-1}$. For the larger
scales, the third term in the Eq.(\ref{zeta}) tends to zero and we
get smaller difference between the density contrast in these two
solutions. In order to compare these results with data, more
detailed simulation in the nonlinear regime of the structure
formation is essential.

%


\section{Age of Universe}
\label{sec6} The age of the universe integrated from the big bang up
to now for a flat universe in terms of free parameters of model $n$
and $X_0$ is given by:
\begin{eqnarray}\label{age55}
t_0(X_p) &=& \int_0^{t_0}\,dt =\int_0^\infty {dz\over (1+z) H(z)}\nonumber\\
&=&\frac{1}{3H_0}\int_{X_p}^{\infty}\frac{F'-XF''}{2F-XF'}\frac{dX}{{\cal{H}}(X)}
\end{eqnarray}

Figure~(\ref{age2}) shows the dependence of $H_0t_0$ (Hubble
parameter times the age of universe) on $X_0$ for a flat universe.
In the lower panel we show the same function for $\Lambda$CDM
universe in terms of $\Omega_\Lambda$ for comparison. As we
expected, $X_0$ in modified gravity behaves as a dark energy and
increasing it makes a longer age for the universe, in the same
direction as increasing the cosmological constant.

 The "age crisis" is one the main reasons
for the acceleration phase of the universe. The problem is that the
universe's age in the CDM universe is less than the age of old stars
in it. Studies on the old stars \cite{carretta00} suggest an age of
$13^{+4}_{-2}$ Gyr for the universe. Richer et. al. \cite{richer02}
and Hasen et. al. \cite{hansen02} also proposed an age of
$12.7\pm0.7$ Gyr, using the white dwarf cooling sequence method.

We use the age of universe in this model for the consistency test
and compare the age of universe with the age of old stars and old
high redshift galaxies (OHRG) in various redshifts. Table \ref{tab2}
shows that the age of universe from the combined analysis of
SNIa$+$CMB$+$SDSS$+$LSS is $14.69_{-0.28}^{+0.29}$ Gyr and
$13.45_{-0.28}^{+0.30}$ Gyr for new gold sample and union data
sample, respectively. These values are in agreement with the age of
old stars \cite{carretta00}. Here we take three OHRG for comparison
with the modified gravity model considering the best fit parameters,
namely the LBDS $53$W$091$, a $3.5$-Gyr old radio galaxy at $z=1.55$
\cite{dunlop96}, the LBDS $53$W$069$ a $4.0$-Gyr old radio galaxy at
$z=1.43$ \cite{dunlop99} and a quasar, APM $08279+5255$ at $z=3.91$
with an age of $t=2.1_{-0.1}^{+0.9}$Gyr \cite{hasinger02}.
To quantify the age-consistency test we introduce the
expression $\tau$ as:
\begin{equation}
 \tau=\frac{t(z;X_0)}{t_{obs}} = \frac{t(z;X_0)H_0}{t_{obs}H_0},
\end{equation}
where $t(z)$ is the age of universe, obtained from the
Eq.(\ref{age55}) and $t_{obs}$ is an estimation for the age of an
old cosmological object. In order to have a compatible age for the
universe we should have $\tau>1$. Table \ref{tab3} reports the value
of $\tau$ for three mentioned OHRGs with various observations. We
see that $f(R)$ modified gravity with the parameters from the
combined observations, provides a compatible age for the universe,
compared to the age of old objects, while  the SNLS data result in a
shorter age for the universe. Once again, APM $08279+5255$ at
$z=3.91$ has a longer age than the universe but gives better results
than some of modified gravity models \cite{sa1}.

\begin{figure}
\epsfxsize=8.5truecm
\begin{center}
\epsfbox{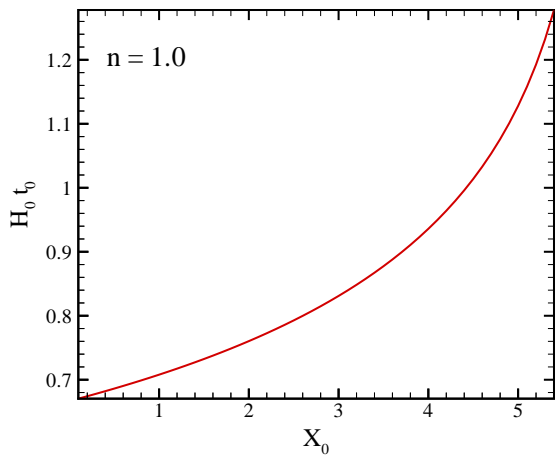}\narrowtext\epsfxsize=8.5truecm
\epsfbox{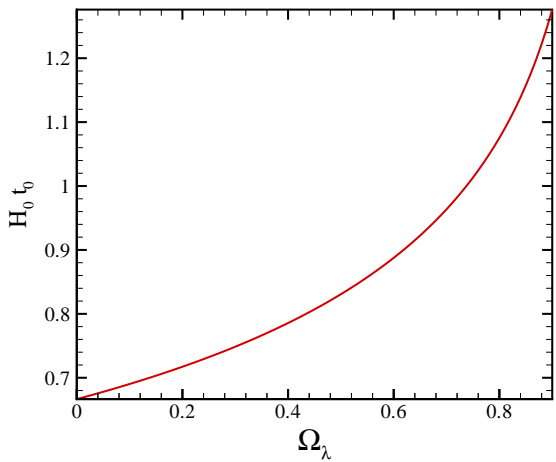} \narrowtext \caption{ $H_0t_0$ (age of
universe times the Hubble constant at the present time) as a
function of $X_0$ (upper panel). $H_0t_0$ for $\Lambda$CDM versus
$\Omega_{ \lambda}$ (lower panel). Increasing $X_0$ gives a longer
age for the universe. This behavior is the same as direction as in
$\Lambda$CDM universe.} \label{age2}
\end{center}
\end{figure}

\begin{figure}
\epsfxsize=8.5truecm
\begin{center}
\epsfbox{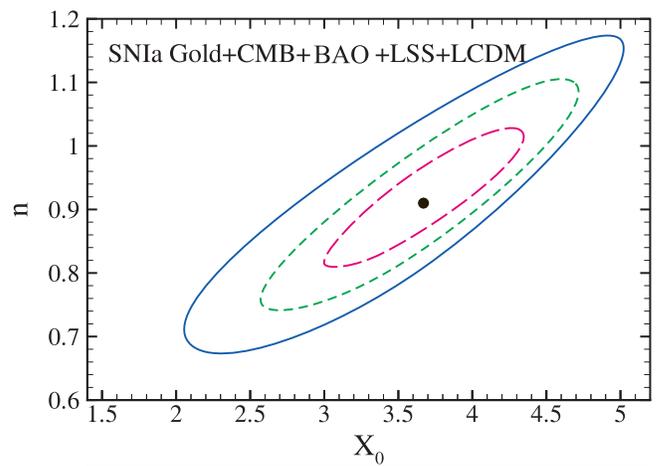}\narrowtext\epsfxsize=8.5truecm
\epsfbox{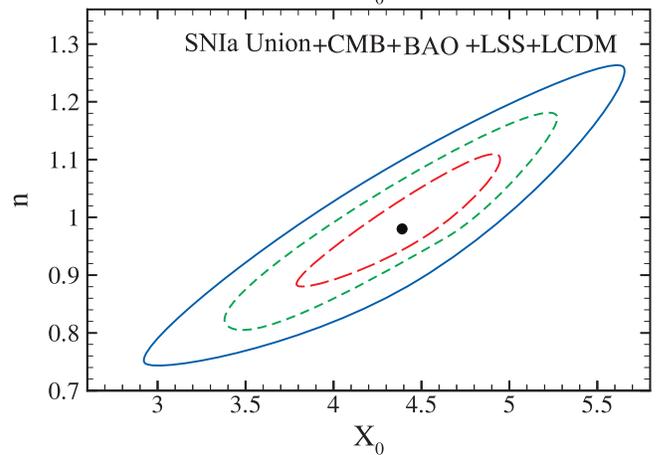} \narrowtext \caption{ Joint likelihood
function in terms of $n$ and $X_0$ considering all the observable
data. In the upper panel the SNIa data is taken from gold sample and
in the lower panel the data is Union sample.} \label{2d}
\end{center}
\end{figure}

\section{conclusion}
In this work we proposed the action of $f(R) = (R^n - R_0^n)^{1/n}$
to obtain the dynamics of the universe. We used the Palatini
formalism to extract the field equation. The advantage of this
formalism is that the field equation is a second-order differential
equation and in the solar system scales we can recover a
Schwarzschild-de-Sitter space with an effective cosmological
constant compatible with the observations. The other advantage of
the Palatini formalism is that it does not suffer from the curvature
instability as pointed out in \cite{sot07}.

We used cosmological tests based on background dynamics such as
Supernova Type Ia, CMB-shift parameter, baryonic acoustic
oscillation and gas mass fraction of the cluster of galaxies. We
also used data from the structure formation to put constrains on the
parameters of the model. Table \ref{tab2} represents constrains on
the parameters of model considering the observational data and their
combination. We also showed that this model provides an age for the
universe sufficiently longer than the age of old astrophysical
objects.

Comparing this model with the observations we put the constrain of
$n = 0.98^{+0.08}_{-0.08}$ for the exponent of action and $X_0 =
4.39^{+0.38}_{-0.38}$ or equivalently $\Omega_m =
0.25^{+0.10}_{-0.10}$. The best value for this model shows that a
standard $\Lambda$CDM model also reside in this range of solution.
Our result is in agreement with the recent work by Kowalski et al.
(2008) where they also obtained almost a $\Lambda$CDM universe with
a nearly constant equation of state for a dark energy model
\cite{kaw08}.

\begin{table}
\begin{center}
\caption{\label{tab2} The best values for the parameters of modified
gravity with the corresponding age for the universe from fitting
with SNIa from new Gold sample and Union data sample, SNIa+CMB,
SNIa+CMB+BAO , SNIa+CMB+BAO+LSS and
SNIa+CMB+BAO+LSS+GAS($\Lambda$CDM) experiments at one and two
$\sigma$ confidence level. The value of $\Omega_m$ is determined
according to equation (\ref{omega})}
\begin{tabular}{|c|c|c|c|c|}
Observation & $n$ &$X_{0}$ & $\Omega_{m}$ & Age (Gyr)
\\ \hline
  &&&& \\
 & $2.01^{+0.72}_{-0.67}$& $6.45_{-1.51}^{+1.13}$&$0.67^{+0.64}_{-0.65}$&$12.99^{+4.42}_{-6.24}$\\ 
 SNIa(new Gold)&&&&\\
 & $2.01^{+1.06}_{-0.95}$& $6.45_{-2.36}^{+1.53}$&$0.67^{+0.90}_{-1.01}$&$12.99^{+7.17}_{-7.17}$
\\ &&&&\\ \hline
&&&&\\
SNIa(new Gold)+& $0.91^{+0.15}_{-0.11}$&$3.58_{-0.58}^{+0.64}$&$0.31^{+0.19}_{-0.14} $ &$14.55^{+2.02}_{-1.68}$ \\
&&&&\\
CMB &$0.91^{+0.23}_{-0.16}$&$3.58_{-0.72}^{+0.92}$&$0.31^{+0.28}_{-0.20}$&$14.55^{+3.40}_{-2.53}$ \\
 &&&&\\ \hline
 &&&&\\
SNIa(new Gold)+& $0.93^{+0.16}_{-0.12}$&$3.76_{-0.59}^{+0.63}$& $0.30^{+0.19}_{-0.15}  $&$14.72^{+2.14}_{-1.89}$ \\
&&&&\\
CMB+BAO &$0.93^{+0.24}_{-0.22}$&$3.76_{-0.83}^{+0.91}$&      $0.30^{+0.29}_{-0.25} $& $14.72^{+3.29}_{-1.43}$\\
&&&&\\ \hline
&&&& \\
SNIa(new Gold)+&
$0.91^{+0.08}_{-0.07}$&$3.67_{-0.42}^{+0.44}$&$0.29^{+0.10}_{-0.09}$&$14.73^{+1.31}_{-1.51}$
 \\
 &&&&\\
 CMB+SDSS+LSS&$0.91^{+0.11}_{-0.10}$&$3.67_{-0.61}^{+0.61}$&$0.29^{+0.15}_{-0.14}$& $14.73^{+1.87}_{-1.68}$ \\
&&&&\\  
 \hline &&&&\\

SNIa(new Gold)+&
$0.91^{+0.08}_{-0.07}$&$3.67_{-0.42}^{+0.44}$&$0.29^{+0.10}_{-0.09}$&$14.73^{+1.31}_{-1.51}$
 \\
 &&&&\\
 CMB+SDSS+LSS+&$0.91^{+0.11}_{-0.10}$&$3.67_{-0.61}^{+0.61}$&$0.29^{+0.15}_{-0.14}$& $14.73^{+1.87}_{-1.68}$ \\
 &&&&\\
 GAS($\Lambda$CDM)&&&&
 \\ \hline
  &&&& \\

  & $1.63^{+0.76}_{-0.92}$&$6.09_{-2.86}^{+1.32}$&$0.47^{+0.69}_{-0.47}$ &$14.27^{+7.31}_{-7.31}$ \\ 
 SNIa (UNION )&&&&\\
 & $1.63^{+1.09}_{-1.10}$&$6.09_{-3.95}^{+1.75}$&$0.47^{+0.90}_{-0.47} $&$14.27^{+8.34}_{-8.34}$
\\ &&&&\\ \hline
&&&&\\
SNIa(UNION)+& $0.99^{+0.15}_{-0.13}$&$4.36_{-0.58}^{+0.57}$&  $0.26_{-0.15}^{+0.17}$ &$15.47^{+2.40}_{-2.43}$ \\
&&&&\\
CMB &$0.99^{+0.22}_{-0.17}$&$4.36_{-0.80}^{+0.83}$&$0.26_{-0.20}^{+0.25}$&$15.47^{+3.77}_{-3.50}$ \\
&&&&\\ \hline
&&&&\\
SNIa(UNION)+& $1.00^{+0.16}_{-0.12}$&$4.45_{-0.54}^{+0.60}$&   $0.26^{+0.18}_{-0.14}  $&$15.59^{+2.64}_{-2.28}$ \\
&&&&\\
CMB+BAO &$1.00^{+0.24}_{-0.17}$&$4.45_{-0.77}^{+0.86}$&   $0.26^{+0.27}_{-0.20}$&$15.59^{+4.17}_{-3.59}  $ \\
&&&&\\ \hline
&&&& \\

SNIa(UNION)+&
$0.98^{+0.08}_{-0.08}$&$4.39_{-0.42}^{+0.38}$&  $0.25_{-0.10}^{+0.10}$&$15.67^{+1.52}_{-1.57} $\\
&&&& \\
 CMB+BAO+LSS&$0.98^{+0.11}_{-0.11}$&$4.39_{-0.60}^{+0.55}$& $0.25_{-0.14}^{+0.14}$& $15.67_{-2.25}^{+2.27} $ \\
&&&&\\ \hline
&&&& \\

SNIa(UNION)+&
$0.98^{+0.08}_{-0.08}$&$4.39_{-0.42}^{+0.38}$&  $0.25_{-0.10}^{+0.10}$&$15.67^{+1.52}_{-1.57}$\\
&&&& \\
 CMB+BAO+LSS&$0.98^{+0.11}_{-0.11}$&$4.39_{-0.60}^{+0.55}$& $0.25_{-0.14}^{+0.14}$& $15.67_{-2.25}^{+2.27}$ \\
&&&&\\
GAS($\Lambda$CDM) &&&& \\ 
\end{tabular}
\end{center}
\end{table}

\begin{table}[htp]
\begin{center}
\caption{\label{tab3} The value of $\tau$ for three high redshift
objects, using the parameters of the model derived from fitting with
the observations at one and two $\sigma$ level of confidences.}

\begin{tabular}{|c|c|c|c|}
  Observation & LBDS &LBDS  & APM  \\
& $53$W$069$&$53$W$091$& $08279+5255$ \\
  & $z=1.43$&$z=1.55$& $z=3.91$  \\ \hline
  &&&\\
SNIa (new Gold)& $0.81^{+0.48}_{-0.81}$ & $0.90^{+0.52}_{-0.90}$& $0.53^{+0.36}_{-0.53}$ \\
&&&\\
&$0.81^{+0.81}_{-0.81}$ & $0.90^{+0.90}_{-0.90}$&$0.53^{+0.53}_{-0.53}$\\
&&&\\
 \hline

&&&\\
SNIa(new Gold)+CMB &$1.18^{+0.29}_{-0.25}$&$1.26^{+0.31}_{-0.28}$&$0.81^{+0.23}_{-0.41}$ \\
 & && \\

&$1.18^{+0.45}_{-0.39}$&$1.26^{+0.49}_{-0.43}$&$0.81^{+0.34}_{-0.49}$\\
&&& \\\hline

&&&\\
SNIa(new Gold)+CMB &$1.20^{+0.31}_{-0.29}$&$1.28^{+0.34}_{-0.31}$&$0.82^{+0.42}_{-0.24}$ \\
 +BAO& && \\
&$1.20^{+0.48}_{-0.46}$&$1.28^{+0.52}_{-0.48}$&$0.82^{+0.50}_{-0.28}$\\
&&&\\\hline
&&&\\
SNIa(new Gold)+CMB & $1.21^{+0.19}_{-0.17}$&$1.29^{+0.20}_{-0.18}$&$0.83^{+0.38}_{-0.14}$ \\
 +BAO+LSS& && \\
 & $1.21^{+0.27}_{-0.25}$&$1.29^{+0.29}_{-0.27}$&$0.83^{+0.43}_{-0.41}$\\
 &&&\\ \hline
&&&\\
SNIa(new Gold)+CMB & $1.21^{+0.19}_{-0.17}$&$1.29^{+0.20}_{-0.18}$&$0.83^{+0.38}_{-0.14}$ \\
 +BAO+LSS+GAS& && \\
 & $1.21^{+0.27}_{-0.25}$&$1.29^{+0.29}_{-0.27}$&$0.83^{+0.43}_{-0.41}$\\
 &&&\\ \hline
&&&\\

 SNIa (UNION)& $ 0.97^{+1.06}_{-0.97}$ & $1.02^{+1.13}_{-1.02}$& $0.63^{+0.73}_{-0.63}$ \\ &&&\\
 &$ 0.97^{+1.15}_{-0.97}$ & $1.02^{+1.49}_{-1.02}$& $0.63^{+0.83}_{-0.63}$ \\
 &&&\\ \hline
&&&\\
SNIa(UNION)+CMB & $1.23^{+0.34}_{-0.36}$&$1.31^{+0.37}_{-0.40}$&$0.84^{+0.44}_{-0.30}$ \\
 & && \\
&
$1.23^{+0.55}_{-0.54}$&$1.31^{+0.60}_{-0.58}$&$0.84^{+0.55}_{-0.45}$\\
&&&\\
 \hline
&&&\\
SNIa(UNION)+CMB & $1.24^{+0.38}_{-0.34}$&$1.33^{+0.41}_{-0.37}$&$0.84^{+0.46}_{-0.28}$ \\
 +BAO& && \\
&
$1.24^{+0.62}_{-0.55}$&$1.33^{+0.67}_{-0.60}$&$0.84^{+0.60}_{-0.47}$\\
&&&\\\hline
&&&\\
SNIa(UNION)+CMB & $1.26^{+0.22}_{-0.23}$&$1.34^{+0.24}_{-0.25}$&$0.86^{+0.40}_{-0.19}$ \\
 +BAO+LSS& && \\
&
$1.26^{+0.33}_{-0.33}$&$1.34^{+0.36}_{-0.36}$&$0.86^{+0.44}_{-0.27}$\\
&&&\\\hline
&&&\\

SNIa(UNION)+CMB & $1.26^{+0.22}_{-0.23}$&$1.34^{+0.24}_{-0.25}$&$0.86^{+0.40}_{-0.19}$ \\
 +BAO+LSS+GAS& && \\
&
$1.26^{+0.33}_{-0.33}$&$1.34^{+0.36}_{-0.36}$&$0.86^{+0.44}_{-0.27}$\\
&&&\\

\end{tabular}
\end{center}
\end{table}


\begin{thebibliography} {50}

\bibitem{lambda}
T. M. Davis et al. Astrophys. J. 666, 716 (2007); E. L. Wright.
Astrophys. J. 664, 633 (2007); M. Sahlen, A. R. Liddle, and D.
Parkinson, Phys. Rev. D 75, 023502 (2007).



\bibitem{modgrv} S. M. Carroll, V. Duvvuri, M. Trodden, and M. S. Turner, Phys.
Rev. D 70, 043528 (2004); S. Nojiri and S. D. Odintsov, Gen.
Relativ. Gravit. 36, 1765 (2004); M. E. Soussa and R. P. Woodard,
Gen. Relativ. Gravit. 36, 855 (2004); G. Allemandi, A. Borowiec, and
M. Francaviglia, Phys. Rev. D 70, 103503 (2004); D. A. Easson, Int.
J. Mod. Phys. A 19, 5343 (2004); S. M. Carroll, A. De Felice, V.
Duvvuri, D. A. Easson, M. Trodden, and M. S. Turner, Phys. Rev. D
71, 063513 (2005); S. Carloni, P. K. S. Dunsby, S. Capozziello, and
A. Troisi, Classical Quantum Gravity 22, 4839 (2005); S.
Capozziello, V. F. Cardone, and A. Troisi, Phys. Rev. D 71, 043503
(2005); G. Cognola, E. Elizalde, S. Nojiri, S. D. Odintsov, and S.
Zerbini, J. Cosmol. Astropart. Phys. 02 (2005) 010; S. Nojiri, S. D.
Odintsov, and S. Tsujikawa, Phys. Rev. D 71, 063004 (2005); T.
Clifton and J. D. Barrow, Phys. Rev. D 72, 103005 (2005); S. Das, N.
Banerjee, and N. Dadhich, Classical Quantum Gravity 23, 4159 (2006);
S. Capozziello, V. F. Cardone, E. Elizalde, S. Nojiri, and S. D.
Odintsov, Phys. Rev. D 73, 043512 (2006); T. P. Sotiriou, Classical
Quantum Gravity 23, 5117 (2006); A. De Felice, M. Hindmarsh, and M.
Trodden, J. Cosmol. Astropart. Phys. 08 (2006) 005; S. Nojiri and S.
D. Odintsov, Phys. Rev. D 74, 086005 (2006); A. F. Zakharov, A. A.
Nucita, F. De Paolis, and G. Ingrosso, Phys. Rev. D 74, 107101
(2006); P. Zhang, Phys. Rev. D 73, 123504 (2006); K. Atazadeh and H.
R. Sepangi, Int. J. Mod. Phys. D 16, 687 (2007); S. M. Carroll, I.
Sawicki, A. Silvestri, and M. Trodden, New J. Phys. 8, 323 (2006);
D. Huterer and E.V. Linder, Phys. Rev. D 75, 023519 (2007);  V.
Faraoni, Phys. Rev. D 74, 104017 (2006); Y. S. Song, W. Hu, and I.
Sawicki, Phys. Rev. D 75, 044004 (2007);  R. Bean, D. Bernat, L.
Pogosian, A. Silvestri, and M. Trodden, Phys. Rev. D 75, 064020
(2007); T. Chiba, T. L. Smith, and A. L. Erickcek, Phys. Rev. D 75,
124014 (2007); V. Faraoni and S. Nadeau, Phys. Rev. D 75, 023501
(2007);  S. Rahvar and Y. Sobouti, Mod. Phys. Lett. A. 23, 1929
(2008).


\bibitem{gr-qc/0505128}
T. Koivisto, Class. Quant. Grav. 23, 4289 (2006).




\bibitem{bagh07}
Sh. Baghram, M. Farhang and S. Rahvar, Phys Rev {\bf D} 75, 044024
(2007).

\bibitem{bagh07b}
M. S. Movahed, Sh. Baghram and S. Rahvar, Phys. Rev. D {\bf {76}},
044008 (2007).

\bibitem{bbn}
C. J. Copi, D. N. Schramm, M. S. Turner, Science 267, 192 (1995).

\bibitem{hst}
W. L. Freedman et al., Astrophys. J. Lett. 553, 47 (2001)

\bibitem{zang}
X. Zhang and F. Q. Wu, Phys. Rev. D 72, 043524 (2005).

\bibitem{perl99}
S. Perlmutter, M. S. Turner, and M. White, Phys. Rev. Lett. 83, 670
(1999).

\bibitem{sch98}
B. P. Schmidt et al., Astrophys. J. 507, 46 (1998).



\bibitem{Riess04}
A. G. Riess  et al., Astrophys. J.{\bf 607}, 655 (2004).

\bibitem{Riess07}
A. G. Riess  et al., Astrophys. J.{\bf 659}, 98 (2007).


\bibitem{Hu97}
W. Hu, N. Sugiyama and J. Silk, Nature {\bf{386}}, 37 (1997).

\bibitem{Hu95}
W. Hu and N. Sugiyama, Astrophys. J. {\bf 444}, 489 (1995).

\bibitem{hu00}
W. Hu, M. Fukugita, M. Zaldarriaga and M. Tegmark, Astrophys. J.
{\bf 549}, 669 (2001).

\bibitem{spe03}
D. N. Spergel, L. Verde, H. V. Peiris et al., Astrophys.
J.{\bf{148}}, 175 (2003).


\bibitem{doran}
M. Doran, M. Lilley, J. Schwindt, and C. Wetterich, Astrophys. J.
559, 501 (2001).


\bibitem{Eis05}
D. J. Eisenstein  et al.,Astrophys. J. {\bf{633}}, 560 (2005).




\bibitem{Perc08}
W. J. Percival et al. MNRAS 381, 1053 (2007).





\bibitem{Sasaki96}
S. Sasaki, PASJ 48,L119 (1996)


\bibitem{pen97}
U. Pen, New Astronomy 2, 309 (1997).

\bibitem{Allen04}
D. Rapetti, S.~W. Allen, and A. Mantz, MNRAS 388, 1265 (2008)





\bibitem{white93}
S. D. M. White, J. F. Navarro, A. E. Evrard and C. S. Frenk, Nature
366, 429 (1993).



\bibitem{Fuk98}
M. Fukugita,C. J. Hogan and P. J. E. Peebles, ApJ 503, 518 (1998).

\bibitem{all04}
S. W. Allen et al., MNRAS 353, 457 (2004).


\bibitem{tavakkol}
K. Uddin, J. E Lidsey and R. Tavakol, Class. Quantum Grav. 24, 3951
(2007).


\bibitem{ver1}
L. Verde, M. Kamionkowski, J. J. Mohr, and A. J. Benson, Mon.
Not. R. Astron. Soc. 321, L7 (2001).

\bibitem{ver2}
O. Lahav, S. L. Bridle, and W. J. Percival (2dFGRS Team), Mon.
Not. R. Astron. Soc. 333, 961 (2002).


\bibitem{koivi06}
Tomi Koivisto and Hannu Kurki-Suonio,Class.Quantum
Grav.{\bf{23}}(2006)2355-2369




\bibitem{carretta00}
E. Carretta et al., Astrophys. J. 533, 215 (2000); B. Chaboyer and
L. M. Krauss, Astrophys. J. Lett. 567, L45 (2002).

\bibitem{richer02}
H. B. Richer et al., Astrophys. J. 574, L151 (2002)

\bibitem{hansen02}
B. M. S. Hansen et al., Astrophys. J. 574, L155 (2002).

\bibitem{dunlop96}
J. Dunlop et al., Nature 381, 581 (1996); H. Spinrard, Astrophys. J.
484, 581 (1997).


\bibitem{dunlop99}
J. Dunlop, in The Most Distant Radio Galaxies, edited by H. J. A.
Rottgering, P. Best, and M. D. Lehnert (Kluwer, Dordrecht, 1999), p.
71.

\bibitem{hasinger02}
G. Hasinger, N. Schartel, and S. Komossa, Astrophys. J. Lett. 573,
L77 (2002).

\bibitem{sa1}
M. S. Movahed and S. Rahvar, Phys. Rev. D 73, 083518 (2006); S.
Rahvar and M. S. Movahed, Phys. Rev. D 75, 023512 (2007).

\bibitem{sot07}
T. P. Sotiriou, Phys. Lett. B 645, 389 (2007).

\bibitem{kaw08}
M. Kowalski et al., Astrophys.J. 686, 749 (2008).




%



\end{thebibliography}
\end{document}